\documentclass[12pt]{article}

\usepackage{amssymb,amsmath,bm,comment}
\allowdisplaybreaks
\renewcommand{\thefootnote}{\fnsymbol{footnote}}

\begin{document}
\title{}

\title{
\begin{flushright}
\begin{minipage}{0.2\linewidth}
\normalsize
WU-HEP-16-18\\
EPHOU-16-015\\*[50pt]
\end{minipage}
\end{flushright}
{\Large \bf 
% Fayet-Iliopoulos terms and right-handed sneutrinos 
% in magnetized orbifold models\\*[20pt] } }
Supersymmetric models on magnetized orbifolds 
with flux-induced Fayet-Iliopoulos terms\\*[20pt] } }

\author{Hiroyuki~Abe$^{1,}$\footnote{
E-mail address: abe@waseda.jp}, \ 
Tatsuo~Kobayashi$^{2,}$\footnote{
E-mail address: kobayashi@particle.sci.hokudai.ac.jp}, \ 
Keigo~Sumita$^{1,}$\footnote{
E-mail address: k.sumita@aoni.waseda.jp}, \ and \ 
Yoshiyuki~Tatsuta$^{1,}$\footnote{
E-mail address: y\_tatsuta@akane.waseda.jp
}\\*[20pt]
$^1${\it \normalsize 
Department of Physics, Waseda University, 
Tokyo 169-8555, Japan} \\
$^2${\it \normalsize 
Department of Physics, Hokkaido University, 
Sapporo 060-0810, Japan} \\*[50pt]}

\date{
\centerline{\small \bf Abstract}
\begin{minipage}{0.9\linewidth}
\medskip 
\medskip 
\small 
We study supersymmetric (SUSY) models derived from the ten-dimensional SUSY 
Yang-Mills theory compactified on magnetized orbifolds, with nonvanishing 
Fayet-Iliopoulos (FI) terms induced by magnetic fluxes in extra dimensions. 
Allowing the presence of FI-terms relaxes a constraint on flux configurations 
in SUSY model building based on magnetized backgrounds. 
In this case, charged fields develop their vacuum expectation values (VEVs) 
to cancel the FI-terms in the D-flat directions of fluxed gauge symmetries, 
which break the gauge symmetries and lead to a SUSY vacuum. 
Based on this idea, we propose a new class of SUSY magnetized orbifold models 
with three generations of quarks and leptons. 
Especially, we construct a model where the right-handed sneutrinos develop their 
VEVs which restore the supersymmetry but yield lepton number violating terms 
below the compactification scale, and show their phenomenological consequences. 
\end{minipage}}

\begin{titlepage}
\maketitle
\thispagestyle{empty}
\clearpage
\tableofcontents
\thispagestyle{empty}
\end{titlepage}
%このコメントアウトとればいつもの
\renewcommand{\thefootnote}{\arabic{footnote}}
\setcounter{footnote}{0}

\section{Introduction} 
There remain several puzzles in the standard model (SM) even though 
the last missing piece, the Higgs boson, was discovered at the 
Large Hadron Collider~\cite{Aad:2012tfa}. 
In particular, an origin of the flavor structure is remarkable one 
in particle physics, that is, the reason why our world consists of 
three generations of the quarks and the leptons, what is more, 
with hierarchical Yukawa couplings. 

The extra dimensions of space are known as a great candidate for the 
new physics beyond the SM, which provides a source of hierarchy~\cite{ArkaniHamed:1999dc} 
for explaining the flavor structure and it has been studied actively so far. 
This is also preferable from a theoretical point of view because 
superstring theories, candidates for the unified theory, 
are defined in ten-dimensional (10D) spacetime. 
We usually consider ten- or other higher-dimensional 
supersymmetric Yang-Mills (SYM) theory to study compactifications of 
the extra dimensional space, because they provide simple frameworks 
for such a purpose and, even more, well motivated by superstring theories. 
We expect that a nontrivial structure in the extra compact space generates 
the observed flavor structures of the SM and realistic models would 
be obtained as four-dimensional (4D) effective field theories of the 
higher-dimensional SYM theories. 

It is known that magnetic fluxes in the compact space have a potential 
to realize the hierarchical flavor structures~\cite{Bachas:1995ik,Cremades:2004wa}. 
Furthermore, in higher-dimensional SYM theories compactified on a product 
of two-dimensional (2D) tori with magnetic fluxes, an analytic form of 
zero-mode wavefunctions can be obtained by solving the Dirac equation, 
which lead to an explicit form of 4D effective Yukawa couplings \cite{Cremades:2004wa}. 
In accordance with this result, a concrete model consistent with the minimal 
supersymmetric SM (MSSM) was constructed with a (semi-)realistic pattern of the 
masses and the mixing angles of quarks and leptons~\cite{Abe:2012fj,Abe:2014soa}. 

We can utilize $Z_N$ orbifolding for constructing more realistic models well 
in the magnetized toroidal compactifications. While the three-generation structure 
is uniquely given with a $\Delta (27)$ flavor symmetry on magnetized tori without 
orbifolding, the orbifold projections as well as certain classes of Wilson-lines 
lead to a broad variety of three-generation structure with different types of 
flavor symmetries~\cite{Abe:2009vi}, and furthermore it can eliminate some 
phenomenologically disfavored extra massless field contents. 
The three-generation structures were systematically studied with 
$Z_2$~orbifolds~\cite{Abe:2008fi,Abe:2008sx} and 
$Z_{3,4,6}$~ones \cite{Abe:2014noa}. 

It is not straightforward to find a supersymmetric (SUSY) SM vacuum on 
magnetized backgrounds, because magnetic flux in extra compact space 
generically produces the Fayet-Iliopoulos (FI) term~\cite{Fayet:1974jb} 
for the hypercharge and/or extra $U(1)$ factors. When there are charged fields 
without their mass terms in the superpotential, the FI-term makes the charged 
fields develop nonvanishing Vacuum Expectation Values (VEVs), which break the 
fluxed gauge symmetry and lead to sizable D-term contributions to the 
charged scalar masses. Thus, in previous works~\cite{Abe:2012fj,Abe:2013bba}, 
it has been required that the FI-terms produced on three 2D tori  
are canceled out by each other in the SUSY model building based on magnetized 
SYM theories. 

In the present paper, we allow configurations of magnetic fluxes which induce 
nonvanishing FI-terms for extra $U(1)$ gauge symmetries other than the $U(1)$ 
hypercharge to construct a new class of models. An anomalous $U(1)$ symmetry 
with a nonvanishing FI-term has been intensively studied in generic 4D SUSY 
models, motivated by string models~\cite{Green:1984sg}. 
In such works, they found that the presence of FI-terms can play a significant 
role in phenomenology of particle physics and cosmology. 
This motivates us to study the magnetized orbifold models with FI-terms 
classically produced by the magnetic fluxes and we expect that the models 
with FI-terms lead to phenomenological implications very different from 
those without the FI-terms. 

In the presence of FI-term in a $U(1)$ sector, charged scalars tend to 
develop their VEVs for the D-term to vanish. In the magnetized models, 
the typical scale of the VEVs is almost equal to the compactification scale. 
Since that is usually as high as the Grand Unification (GUT) scale or 
Planck scale, the responsible fields must be a SM singlet. 
In this paper, we especially focus on the right-handed sneutrinos, 
which are SM singlets but can play phenomenologically relevant roles 
in the MSSM sector, and consider their nonvanishing VEVs in the D-flat 
directions of fluxed $U(1)$ symmetries to cancel the FI-term. 

This paper is organized as follows. 

In Section~\ref{sec:FI}, 
we give an overview of magnetized orbifold models and show some 
basic ideas for realizing a new class of models with flux-induced FI-terms. 
We first explain the $D$-terms in the toroidal compactification of 10D SYM 
theories with magnetic fluxes and the orbifolding in Section~\ref{ssec:D}, 
and show how to construct the MSSM-like models by introducing flux-induced 
FI-terms in some $U(1)$ subgroups of $U(8)$ SYM theory on a magnetized 
orbifold in Section~\ref{ssec:U8}. 

Subsequently, in Section~\ref{sec:model}, we construct a SUSY model with 
flux-induced FI-terms for certain $U(1)$ subgroups of $U(8)$, which make 
the right-handed sneutrinos  develop their VEVs in the D-flat directions 
leading to a SUSY minimum. We show the almost unique flux configuration 
which realizes three generations of quarks and leptons at the SUSY minimum 
where the sneutrinos have VEVs in Section~\ref{ssec:3gen}. 
In Section~\ref{ssec:snv}, 
we show the superpotential of our model and discuss a relation between 
the textures of the $\mu$-terms and the lepton number violating mass terms. 
Their interplay modifies the flavor structure of the leptons. 
We estimate the mass ratios and mixing angles of the quarks and the leptons 
in a numerical calculation in Section~\ref{ssec:hv}, 
where we show that our model leads to a hopeful spectrum of the SM matter fields. 

Finally, we summarize our result and discuss its future prospect 
in Section~\ref{sec:summary}. 

The analytic forms of Yukawa couplings are shown in Appendix~\ref{sec:app} 
for the model shown in Section~\ref{ssec:3gen}.

\section{Flux-induced FI-terms on magnetized orbifolds}
\label{sec:FI}
We briefly review the magnetized compactification with/without 
orbifolding in 10D SYM theories. 
It will be shown that the magnetic 
fluxes break the gauge symmetry down to a product of several unbroken 
gauge subgroups and bifundamental gaugino fields of the unbroken subgroups 
have degenerate zero-modes with their conjugate ones eliminated, 
that is, generations of chiral fermions like the SM are obtained. 

The magnetic fluxes in general produce FI-terms for some of the 
unbroken gauge subgroups. In previous works~\cite{Abe:2012fj,Abe:2013bba} 
for SUSY model building, it has been required 
that the FI-terms do not appear for any of the subgroups, otherwise 
SUSY is broken or those can lead to color and/or electromagnetism 
breaking vacua in some cases. 
The conditions for the FI-terms to vanish is so strong that we have 
been found a few configurations of magnetic fluxes which lead to 
three generations of the quarks and leptons preserving SUSY 
so far~\cite{Abe:2013bba}. 

In the present paper, we consider model building such  that the flux-induced 
FI-terms are nonvanishing for some $U(1)$s (other than the hypercharge 
$U(1)$) and the charged (but SM-singlet) scalar fields develop their 
VEVs to cancel the FI-terms in the D-term potential, 
leading to a SUSY minimum of the scalar potential. 
Then we will see in the following and in the next section that, 
the allowed flux configurations for the SUSY model building 
and the resultant phenomenologies are quite different from 
those with vanishing FI-terms. 

\subsection{$D$-terms on magnetized tori}
\label{ssec:D}
First we review the 10D SYM theories compactified on magnetized tori 
and orbifolds. We exclusively consider a product of three 2D tori 
as a six-dimensional extra compact space, 
which are described by complex coordinates $(z_i,\bar z_i)$ ($i=1,2,3$). 
The 10D SYM theory contains a pair of 10D gauge field $A_M$ ($M=0,1,\ldots,9$) 
and 10D Majorana-Weyl spinor field $\lambda$. 

We decompose them into 
a 4D vector $A_\mu$ ($\mu=0,1,2,3$), 
three 4D complex scalars $\varphi_i$ and 
four 4D Weyl spinor fields $\lambda_{\pm\pm\pm}$ as 
\begin{equation*}
A_M=(A_\mu,\varphi_i),\qquad\lambda = 
(\lambda_{+++},\lambda_{+--},\lambda_{-+-},\lambda_{--+}). 
\end{equation*}
Here, the complex scalar field $\varphi_i$ is defined in accordance with 
the complex coordinate $z_i$. Each subscript ``$\pm$'' accompanied with 
4D Weyl spinors $\lambda_{\pm\pm\pm}$ represents the chirality on one 
of the three 2D tori. 
4D Weyl spinors with other combinations of $\pm$s, e.g., $\lambda_{---}$, 
do not appear because here the 10D spinor field $\lambda$ is an 
eigenstate of a 10D chirality operator with a positive eigenvalue 
as a result of the Majorana-Weyl condition. 

These component fields form 4D $\mathcal N=1$ supermultiplets 
with auxiliary fields $F_i$ and $D$ as 
\begin{eqnarray*}
V&\equiv& -\theta\sigma^\mu\bar\theta A_\mu+i\bar\theta\bar\theta\theta\lambda_0
-i\theta\theta\bar\theta\bar\lambda_0+\frac12\theta\theta\bar\theta\bar\theta D,\\
\phi_i&\equiv& \frac1{\sqrt2} A_i+\sqrt2\theta\lambda_i+\theta\theta F_i,
\end{eqnarray*}
where $(\lambda_0,\lambda_1,\lambda_2,\lambda_3)=
(\lambda_{+++},\lambda_{+--},\lambda_{-+-},\lambda_{--+})$. 
In Ref.~\cite{Marcus:1983wb,ArkaniHamed:2001tb},  the 10D SYM action 
is expressed  in the 4D $\mathcal N=1$ superspace 
by using these superfields. 
We compactify it on a product of the 4D Minkowski spacetime and three 2D tori, 
$M^4\times T^2\times T^2\times T^2$, and introduce Abelian constant magnetic 
fluxes on the tori, which was studied in Ref.~\cite{Abe:2012ya} 
and its 4D effective action is derived in a systematic way shown there. 

Let us consider 10D $U(N)$ SYM theories 
compactified on three 2D tori with magnetic fluxes of $(1,1)$ form, e.g., 
\begin{equation*}
F_{z_i\bar z_{\bar i}}=2\pi M^{(i)}=2\pi
\begin{pmatrix}
M^{(i)}_a\bm1_{N_a}&0\\
0&M^{(i)}_b\bm1_{N_b}
\end{pmatrix},  \qquad M^{(i)}_a, M^{(i)}_b \in \mathbb{Z},
\end{equation*}
in the $U(N)$ gauge space with $N=N_a+N_b$, 
where integer values of  $M^{(i)}_a$ and $M^{(i)}_b$ 
represent the quantized magnetic fluxes 
and $\bm1_{N_a}$ denotes the $(N_a \times N_a)$ unit matrix. 
When the fluxes, $M^{(i)}_a$ and $M^{(i)}_b$, are different from each other, 
the gauge symmetry is broken 
as $U(N)\rightarrow U(N_a)\times U(N_b)$. 
In this paper, we will not consider magnetic fluxes of the other forms, i.e., 
$(2,0)$ and $(0,2)$ forms. 
On this magnetized background, $U(N)$ adjoint field is decomposed as 
\begin{equation}
V=
\begin{pmatrix}
V_{aa}&V_{ab}\\
V_{ba}&V_{bb}
\end{pmatrix}, \qquad 
\phi_i=
\begin{pmatrix}
(\phi_i)_{aa}&(\phi_i)_{ab}\\
(\phi_i)_{ba}&(\phi_i)_{bb}
\end{pmatrix}. 
\label{decomp}
\end{equation}

The VEVs of zero-modes of auxiliary fields $D_a$ and $(F_i)_a$ in 
$U(N_a)$-adjoint vector and chiral superfields $V_{aa}$ and $(\phi_i)_{aa}$ 
are determined by their equations of motion respectively as 
\begin{equation}
D_a= \left( 
\frac1{\mathcal A^{(1)}}M^{(1)}_a+\frac1{\mathcal A^{(2)}}M^{(2)}_a
+\frac1{\mathcal A^{(3)}}M^{(3)}_a \right) \times \bm1_{N_a}, 
\qquad (F_i)_a=0 \times \bm1_{N_a}, 
\label{eq:aux}
\end{equation}
where we denote the $i$-th 2D torus area by $\mathcal A^{(i)}$. 
Then the conditions for preserving 4D $\mathcal N=1$ SUSY 
are simply described by $D_a=0$, i.e., 
\begin{equation}
\frac1{\mathcal A^{(1)}}M^{(1)}_a+\frac1{\mathcal A^{(2)}}M^{(2)}_a
+\frac1{\mathcal A^{(3)}}M^{(3)}_a=0, 
\label{eq:susy}
\end{equation}
as long as we restrict ourselves to the case with 
vanishing VEVs of fields in the bifundamental representations 
$(N_a, \bar N_b)$ and/or $(\bar N_a, N_b)$ carried by 
$(\phi_i)_{ab}$ and/or $(\phi_i)_{ba}$. 
Similar arguments hold for $U(N_b)$-adjoint superfields 
by replacing $a$ with $b$ in Eqs.~(\ref{eq:aux}) and (\ref{eq:susy}). 
These give constraints on magnetic fluxes $M_a^{(i)}$ and $M_b^{(i)}$, 
and the supersymmetric model building on magnetized tori and orbifolds 
suffers from them. 

The bifundamental fields in $(\phi_i)_{ab}$ feel the 
$M^{(i)}_{ab}\equiv M^{(i)}_a-M^{(i)}_b$ unit 
of magnetic fluxes on the $i$-th 2D torus. 
For the positive values of $M^{(i)}_{ab}$, 
the magnetic fluxes give rise to $M^{(i)}_{ab}$ degenerate zero-modes 
for $(\phi_i)_{ab}$ and $(\phi_{j\neq i})_{ba}$, 
and their conjugate ones, $(\phi_{i})_{ba}$ and $(\phi_{j\neq i})_{ab}$, 
are then eliminated in the low-energy spectrum, which yield 
generations of chiral fermions~\cite{Bachas:1995ik,Cremades:2004wa}. 
For the negative values of $M^{(i)}_{ab}$, in contrast, 
$|M^{(i)}_{ab}|$ degenerate zero-modes are produced for 
$(\phi_{i})_{ba}$ and $(\phi_{j\neq i})_{ab}$, 
while $(\phi_{i})_{ab}$ and $(\phi_{j\neq i})_{ba}$ have no zero-modes. 
Note that, with the vanishing value of $M^{(i)}_{ab}$, 
all of representations have a single zero-mode with a flat wavefunction. 

Next, we go on to the compactification on magnetized orbifolds. 
Systematic studies of $Z_N$ ($N$=3,4,6) orbifolds with magnetic fluxes 
have been recently done~\cite{Abe:2014noa}. 
In this paper we concentrate on the $Z_2$ orbifolds~\cite{Abe:2008fi,Abe:2008sx}, 
where all field contents are assigned into either even or odd modes under the $Z_2$ parity. 
The numbers of degenerate even or odd zero-modes produced by 
the magnetic fluxes are reduced because of the orbifold projection and 
these zero-modes numbers are shown 
in Table \ref{tb:MOzeros}~\cite{Abe:2008fi}. 
\begin{table}[t]
\center
\begin{tabular}{ccccccccccccc}
 $M$&$0$& $1$  &$2$  &$3$  &$4$  &$5$  &$6$ & $7$ & $8$ & $9$ & $2n$&$2n+1$ \\\hline
Even& $1$  &$1$  &$2$  &$2$  &$3$  &$3$  &$4$ & $4$ & $5$ & $5$ &$n+1$&$n+1$\\
Odd& $0$  &$0$  &$0$  &$1$  &$1$  &$2$  &$2$ & $3$ & $3$ & $4$ &$n-1$&$n$
\end{tabular}
\caption{The number of degenerate zero-modes on magnetized $Z_2$ orbifolds 
where $n\in\mathbb{N}$~\cite{Abe:2008fi}.}
\label{tb:MOzeros}
\end{table}

It is also remarkable in our discussion that these $Z_2$ parities 
are assigned by respecting the superfield formulation where the 
4D ${\cal N}=1$ SUSY is manifest. This determines the transformation 
low of the superfields under the $Z_2$. For example, under a $Z_2$ 
operation given by $(z_1,z_2,z_3) \rightarrow (-z_1,-z_2,z_3)$, 
they transform as 
\begin{eqnarray*}
V&\rightarrow& +PVP^{-1},\\
\phi_1&\rightarrow& - P\phi_1P^{-1},\\
\phi_2&\rightarrow& - P\phi_2P^{-1},\\
\phi_3&\rightarrow& + P\phi_3P^{-1},
\end{eqnarray*}
where $P$ is a projection operator ($P^2={\bm1}$) 
and is an $N\times N$-matrix in $U(N)$ cases. 

\subsection{$U(8)$ models with FI-terms}
\label{ssec:U8}
In the following, we discuss the FI-term induced by the magnetic fluxes 
in three-generation magnetized orbifold models. 
The Pati-Salam like gauge group $U(4)_C\times U(2)_L\times U(2)_R$ can realize 
a realistic model based on magnetized backgrounds. 
This is derived from the $U(8)$ SYM theory with magnetic fluxes of the form 
\begin{equation}
M^{(i)}=\begin{pmatrix}
M^{(i)}_C\times {\bm 1}_4&0&0\\
0&M^{(i)}_L\times {\bm 1}_2&0\\
0&0&M^{(i)}_R\times {\bm 1}_2, 
\end{pmatrix}, \label{eq:PSflux}
\end{equation}
which breaks $U(8)$ down to $U(4)_C\times U(2)_L\times U(2)_R$. 
The representation $({\bm 1}, {\bm 2}, \bar {\bm2})$ 
of the respective unbroken subgroups contains the Higgs multiplets. 
The left-handed and right-handed matter fields are assigned into 
the representation $({\bm 4}, \bar{\bm 2}, {\bm1})$ and 
$(\bar{\bm 4}, {\bm1}, {\bm2})$, respectively. 
The right-handed sneutrinos (to get nonvanishing VEVs later) 
are contained in the representation $(\bar{\bm 4}, {\bm1}, {\bm2})$. 

These magnetic fluxes generically produce FI-terms for the 
Abelian part of each unbroken gauge subgroup. For example, 
in the $U(4)_C$ sector, the magnetic fluxes on the three 2D tori 
yield constant contributions in the D-term of $U(4)_C$, 
\begin{equation}
D_C=\left(\frac1{\mathcal A^{(1)}}M_C^{(1)}
+\frac1{\mathcal A^{(2)}}M_C^{(2)}
+\frac1{\mathcal A^{(3)}}M_C^{(3)}\right)\times {\bm 1}_4, 
\label{eq:fi} 
\end{equation}
like in Eq.~(\ref{eq:aux}), 
due to the presence of FI-term for the $U(4)_C$ 
vector superfield induced by the fluxes $M_C^{(i)}\times {\bm 1}_4$. 
In this paper, we adopt such flux configurations, which  generate 
nonvanishing FI-terms for the fluxed $U(1)$ vector multiplet, and 
consider the case that some SM singlets develop their VEVs to cancel 
the FI-terms in the 4D effective field theories\footnote{
This situation may be realized by some non-Abelian gauge backgrounds 
in the original higher-dimensional field theory, where certain 
off-diagonal elements of higher-dimensional gauge field have 
constant backgrounds. In this case, however, such off-diagonal 
constants could affect the wavefunction profiles of zero-modes. 
We will not discuss such a case here but consider VEVs in the 
D-flat directions of 4D effective field theory, minimizing the 
4D scalar potential.} restoring the $\mathcal N=1$ SUSY.

Let us consider the D-term of $U(N_a)$ subgroup with 
the VEVs of matter fields in its fundamental representation. 
The coupling of such a matter chiral superfield $\Phi$ 
and the gauge multiplet 
is described by 
\begin{equation*}
\int d^4\theta~(\Phi^*)_i(e^{V})_{ij}(\Phi)_j,
\end{equation*}
where $V$ is the $U(N_a)$ gauge superfield, 
and $i,j=1,2,\ldots,N_a$ are now $U(N_a)$ indices. 
Without the flux-induced FI-term, the D-term of $U(N_a)$ 
is given by $D_{ij}=\langle(\Phi^*)_i\rangle\langle(\Phi)_j\rangle$ 
which can be always diagonalized as 
\begin{equation}
D_{ij}={\rm diag} (x,0,\ldots,0), 
\label{eq:diagd}
\end{equation}
by a certain $U(N_a)$ rotation, where $x$ is a real constant. 
In the presence of the FI-term, a SUSY vacuum is obtained 
when the following condition is satisfied: 
\begin{equation}
\delta_{ij}\left(\frac1{\mathcal A^{(1)}}M_a^{(1)}
+\frac1{\mathcal A^{(2)}}M_a^{(2)}
+\frac1{\mathcal A^{(3)}}M_a^{(3)}\right)
+\langle(\Phi^*)_i\rangle\langle(\Phi)_j\rangle=0. 
\label{eq:undflat}
\end{equation}
For $N_a>1$, this cannot be satisfied because the first contribution 
is rank $N_a$ but the second one is rank $1$ as we see in Eq.~(\ref{eq:diagd}). 
Therefore we find that Eq.~(\ref{eq:undflat}) can be satisfied only 
in the case with $N_a=1$. 

From the above argument we expect that, when the unbroken subgroup 
which has the nonvanishing FI-term is $U(1)$, the FI-term can be 
canceled by the VEVs of charged fields. In this case, the scale of 
VEVs is comparable to the compactification scale, which would be typically 
set to $M_{\rm GUT}$ or $M_{\rm Planck}$. In the Pati-Salam like 
model obtained by the flux configuration~(\ref{eq:PSflux}), such a 
large value of VEV is phenomenologically allowed only for the 
right-handed sneutrinos. We identify them as the responsible field for 
canceling the FI-term. In the following, we adopt the flux 
configurations with which all the unbroken gauge subgroups related 
to the right-handed neutrinos are $U(1)$, and consider the case that 
their flux-induced FI-terms are canceled by the VEVs of right-handed 
sneutrinos, yielding a new class of SUSY vacua. 

In the Pati-Salam like model, the right-handed neutrinos are 
carried by the bifundamental representation of $U(4)_C\times U(2)_R$. 
In accordance with the above discussion, these two gauge groups 
have to be further broken by the magnetic fluxes down to 
$U(3)_C\times U(1)_\ell$ and $U(1)_r\times U(1)_{r'}$ from the beginning. 
This gauge symmetry breaking is realized by the magnetic fluxes of the form, 
\begin{equation}
M^{(i)}=\begin{pmatrix}
M^{(i)}_C\times {\bm 1}_3&0&0&0&0\\
0&M^{(i)}_\ell&0&0&0\\
0&0&M^{(i)}_L\times {\bm 1}_2&0&0\\
0&0&0&M^{(i)}_r&0\\
0&0&0&0&M^{(i)}_{r'}
\end{pmatrix},
\label{eq:5flux}
\end{equation}
where each flux number of 
$M^{(i)}_C$, $M^{(i)}_\ell$, $M^{(i)}_L$, $M^{(i)}_r$ and $M^{(i)}_{r'}$ 
takes a different value from the others on at least one of three 2D tori $i=1,2,3$, 
otherwise the unbroken gauge symmetry is enhanced. 
Note that, this form of the magnetic fluxes can be shifted as 
$M^{(i)}\rightarrow M^{(i)}+\alpha\times{\bm 1}_8$ 
without changing the spectrum in the low-energy effective field theory, 
and in the following we set the value of $M^{(i)}_C$ to vanish 
by using this degree of freedom. 

On this magnetized background, the gauge symmetry is broken as 
$U(8) \rightarrow 
U(3)_C\times U(1)_\ell\times U(2)_L\times U(1)_r\times U(1)_{r'}$. 
We can then assign the MSSM fields into the decomposed adjoint fields as 
\begin{equation}
\Phi_{\rm adj}=
\begin{pmatrix}
*&*&Q&*&*\\
*&*&L&*&*\\
*&*&*&H_u&H_d\\
U&N&*&*&*\\
D&E&*&*&*
\end{pmatrix}.
\label{eq:5fluxfield}
\end{equation}
The substructure of this $8 \times 8$ matrix $\Phi_{\rm adj}$ is 
exactly the same as Eq.~(\ref{eq:5flux}). The fields denoted by 
$Q$, $L$, $U$, $D$, $N$, $E$, $H_u$ and $H_d$ correspond to 
the left-handed quarks, 
the left-handed leptons, 
the right-handed up-type quarks, 
the right-handed down-type quarks, 
the right-handed neutrinos, 
the right-handed charged leptons, 
the up-type Higgs fields 
and the down-type Higgs fields, respectively. 
The other elements symbolically expressed by~$\ast$ represent 
extra fields which can be eliminated by the interplay between the  
magnetic fluxes and orbifold projections. 
The VEVs of the right-handed sneutrinos can give rise to new 
contributions in $U(1)_\ell$ and $U(1)_r$ D-terms. 
The sneutrino VEVs break one linear combination of these two $U(1)'s$ 
while the other (orthogonal) combination is preserved, 
and the latter one is a part of $U(1)$ hypercharge. 

Because it is required for our purpose that each flux number of 
$M^{(i)}_C$, $M^{(i)}_\ell$, $M^{(i)}_L$, $M^{(i)}_r$ and $M^{(i)}_{r'}$ 
in Eq.~(\ref{eq:5flux}) takes a different value from 
the others on at least one of three 2D tori, 
in order to obtain three generations of quarks $Q$, $U$, $D$ 
and leptons $L$, $N$ $E$, certain orbifold projections 
are necessary. Without orbifolding, the three-generation 
structure is generated by $M=3$ unit of magnetic fluxes 
exclusively, and any flux configurations which induce 
three generations of the quarks and leptons cannot 
realize the required pattern of gauge symmetry breaking. 
In contrast, as we find in Table~\ref{tb:MOzeros}, 
three generations appear with $M=4,5,7$ or $8$ unit 
of magnetic fluxes\footnote{
An $M=6$ unit of magnetic flux on twisted orbifolds~\cite{Abe:2013bca} 
can also induce three-generation structure.}, 
which allow us to construct three-generation models 
with the desired patterns of gauge symmetry breaking 
for our purpose. 

If we do not allow any nonvanishing VEVs of off-diagonal 
(bifundamental) fields in Eq.~(\ref{eq:5fluxfield}) in 
our model building, the SUSY preserving conditions~(\ref{eq:susy}) 
for all the unbroken gauge subgroups $a=C$, $\ell$, $L$, $r$ and $r'$ 
severely restrict the patterns of original flux configurations. 
Indeed, SUSY configurations of the magnetic fluxes which 
lead to a product gauge group with more than three subgroups 
have never been found~\cite{Abe:2014vza}. 
On the other hand, if we consider a situation that any 
off-diagonal fields in Eq.~(\ref{eq:5fluxfield}), 
especially the right-handed sneutrinos denoted by $N$ 
from the phenomenological viewpoint in the Pati-Salam like model, 
develop their nonvanishing VEVs, there appear additional 
contributions in the D-flat conditions and the SUSY preserving 
condition~(\ref{eq:susy}) is modified as 
\begin{equation}
\frac1{\mathcal A^{(1)}}M^{(1)}+\frac1{\mathcal A^{(2)}}M^{(2)}
+\frac1{\mathcal A^{(3)}}M^{(3)}+X=0,
\label{eq:modsusy}
\end{equation}
where $M^{(i)}$ is shown in Eq.~(\ref{eq:5flux}), and 
the $(8\times8)$-matrix $X$ represents the contributions due to 
the VEVs $\langle \tilde\nu_i\rangle$ of right-handed sneutrinos. 
Because the right-handed neutrinos are charged under 
$U(1)_\ell$ and $U(1)_r$, the matrix $X$ is described as 
\begin{equation}
X=\begin{pmatrix}
0&0&0&0&0\\
0&qx&0&0&0\\
0&0&0&0&0\\
0&0&0&-qx&0\\
0&0&0&0&0
\end{pmatrix},
\label{eq:neuvev}
\end{equation}
which is parametrized by 
$x=\sum_i\langle\tilde\nu_i\rangle^2$ and $q=\pm 1$. 
The modified SUSY condition~(\ref{eq:modsusy}) 
allows a new class of SUSY vacua on magnetized orbifolds, 
which we demonstrate in the next section.

\section{Supersymmetric models with FI-terms}
\label{sec:model}
Magnetized orbifolds provide a wide variety of three-generation structure. 
One of the key points to construct phenomenological models is that 
three-generation structure for quarks and leptons must be produced 
on a single 2D torus, otherwise the rank of Yukawa matrices is reduced to one. 
For this reason, we concentrate on a 2D torus for a while. 
The model building based on magnetized $Z_2$ orbifolds was 
studied systematically in Ref.~\cite{Abe:2008sx,Abe:2014vza}. 
On (untwisted) magnetized $Z_2$ orbifolds, three generations of 
chiral fermions are produced by the $|M|=4,5$ units of fluxes for $Z_2$ 
even modes, and $|M|=7,8$ for odd modes as shown in Table~\ref{tb:MOzeros}. 
There is a severe constraint, as well as the SUSY preserving condition, 
on the flux configurations due to the requirement that the numbers of 
$H_u$ and $H_d$ have to be equal to each other in order to avoid 
the anomaly of $U(1)$ hypercharges. 

\subsection{The three-generation model}
\label{ssec:3gen}
A systematic search performed in Ref.~\cite{Abe:2014vza} shows that 
there are only four patterns of magnetic fluxes and $Z_2$ parity 
assignments on one of three 2D tori, which would be available for our purpose, 
namely, the gauge symmetry is suitably broken as 
$U(8)\rightarrow 
U(3)_C\times U(1)_\ell\times U(2)_L\times U(1)_r\times U(1)_{r'}$, 
and three generations of quarks and leptons 
as well as pairs of $H_u$ and $H_d$ are obtained. 
One of them adopted in the following analysis is shown 
in Table~\ref{tb:flux}, where five pairs of the Higgs fields appear. 
\begin{table}[t]
\center
\begin{tabular}{|c|rcr|c|c|}\hline
& \# of fluxes & & &$Z_2$ parity& $\#$ of zero-modes\\\hline
$Q$&$M_C^{(i)}-M_L^{(i)}$&$=$&$-4$&even&3\\\hline
$L$&$M_\ell^{(i)}-M_L^{(i)}$&$=$&$-5$&even&3\\\hline
$U$&$M_r^{(i)}-M_C^{(i)}$&$=$&$-5$&even&3\\\hline
$D$&$M_{r'}^{(i)}-M_C^{(i)}$&$=$&$-8$&odd&3\\\hline
$N$&$M_r^{(i)}-M_\ell^{(i)}$&$=$&$-4$&even&3\\\hline
$E$&$M_{r'}^{(i)}-M_\ell^{(i)}$&$=$&$-7$&odd&3\\\hline
$H_u$&$M_L^{(i)}-M_r^{(i)}$&$=$&9&even&5\\\hline
$H_d$&$M_L^{(i)}-M_{r'}^{(i)}$&$=$&12&odd&5\\\hline
\end{tabular}
\caption{An example of magnetic fluxes felt by MSSM 
fields and the parity assignments for them under the 
$Z_2$ projection $z_i \rightarrow -z_i$ is shown.}
\label{tb:flux}
\end{table}
Note that the assignment of $Z_2$ parity in this model is consistent 
with the nonvanishing Yukawa couplings required in (MS)SM. 
Other three patterns are obtained by exchanging the assignment 
of flux and parity between the quark and lepton sectors, 
and/or, the up and down sectors in the above example. 

The flux configurations and the $Z_2$ parity assignment 
on the other two 2D tori are mainly determined in order to satisfy 
the modified version of the SUSY preserving condition (\ref{eq:modsusy}) 
and induce no extra generations of quarks and leptons. 
These two conditions are indeed so severe that 
there exists one and only possible configuration which we find 
in a systematic search. 
That is given by 
\begin{equation}
M^{(1)}=\begin{pmatrix}
0&0&0&0&0\\
0&-1&0&0&0\\
0&0&4&0&0\\
0&0&0&-8&0\\
0&0&0&0&-5
\end{pmatrix},\ 
M^{(2)}=\begin{pmatrix}
0&0&0&0&0\\
0&0&0&0&0\\
0&0&-1&0&0\\
0&0&0&0&0\\
0&0&0&0&-1
\end{pmatrix},\ 
M^{(3)}=\begin{pmatrix}
0&0&0&0&0\\
0&0&0&0&0\\
0&0&0&0&0\\
0&0&0&1&0\\
0&0&0&0&1
\end{pmatrix}, 
\label{eq:magmag}
\end{equation}
and 
\begin{equation*}
X=\begin{pmatrix}
0&0&0&0&0\\
0&1&0&0&0\\
0&0&0&0&0\\
0&0&0&-1&0\\
0&0&0&0&0
\end{pmatrix}.
\end{equation*} 
This matrix $X$ corresponds to Eq.~(\ref{eq:neuvev}) with $q=x=+1$. 
These satisfy the SUSY preserving conditions (\ref{eq:modsusy}) with 
$\mathcal A^{(1)}/\mathcal A^{(2)}=4$ and $\mathcal A^{(1)}/\mathcal A^{(3)}=9$. 
In this case, the sneutrino VEVs are given in the unit of $1/\sqrt{\mathcal A^{(1)}}$, 
which we identify with the compactification scale because 
this is just the mass scale of the first excited Kaluza-Klein mode. 
Note that, exchanging flux configurations on the three 2D tori leads to 
different models on first glance, but they are physically equivalent to each other. 
It is just a matter of labeling the complex coordinates of three 2D tori. 

On this magnetized background, we can obtain the three generations 
of quarks and leptons and the five pairs of Higgs fields 
when the $Z_2$ parities on each 2D torus $(z_i,\bar z_i)$
are assigned as shown in Table~\ref{tb:z2}. 
\begin{table}[t]
\center
\begin{tabular}{|c|c|c|c|c|}\hline
&$(z_1,\bar z_1)$&$(z_2,\bar z_2)$&$(z_3,\bar z_3)$\\\hline
$Q$&$-4$, even&+1, even&0, even\\\hline
$L$&$-5$, even&+1, even&0, even\\\hline
$U$&$-8$, odd&0, even&+1, even\\\hline
$D$&$-5$, even&$-1$, even&+1, even\\\hline
$N$&$-7$, odd&0, even&+1, even\\\hline
$E$&$-4$, even&$-1$, even&+1, even\\\hline
$H_u$&+12, odd&$-1$, even&$-1$, even\\\hline
$H_d$&+9, even&0, even&-1, even\\\hline
\end{tabular}
\caption{We summarize the effective magnetic fluxes felt 
by each of MSSM fields and the suitable $Z_2$ parities.}
\label{tb:z2}
\end{table}
The orbifolding in $(z_2,\bar z_2)$ and $(z_3,\bar z_3)$ directions 
are not necessary to realize the three generations, but 
we impose them on the two 2D tori 
because they are useful to eliminate extra field contents, 
such as phenomenologically disfavored chiral exotics and 
massless adjoint fields. Note that, all the MSSM fields are assigned 
to $Z_2$ even mode on these two 2D tori otherwise they are eliminated 
in the 4D low-energy spectrum. 

We consider $T^6/Z_2\times Z'_2$ orbifold to 
realize the desired parity assignment for our purpose. 
The orbifold projection operators are assigned as mentioned at the 
end of Section~\ref{ssec:D}; $\phi_i\rightarrow \pm P\phi_i P^{-1}$. 
We find that the following ones lead to the desirable $Z_2$ parities, 
\begin{eqnarray}
Z_2&:&(z_1,\,z_2,\,z_3)\rightarrow(-z_1,\,-z_2,\,z_3) 
\qquad {\rm with} \qquad 
P=\begin{pmatrix}
1&0&0&0&0\\
0&1&0&0&0\\
0&0&-1&0&0\\
0&0&0&-1&0\\
0&0&0&0&1
\end{pmatrix},
\nonumber \\
Z'_2&:&(z_1,\,z_2,\,z_3)\rightarrow(z_1,\,-z_2,\,-z_3) 
\qquad {\rm with} \qquad 
P'=\begin{pmatrix}
1&0&0&0&0\\
0&1&0&0&0\\
0&0&-1&0&0\\
0&0&0&-1&0\\
0&0&0&0&-1
\end{pmatrix}. 
\label{eq:po}
\end{eqnarray}
On this magnetized orbifold, chiral superfields $\phi_i$ produce 
the following zero-modes, 
\begin{equation*}
\phi_1=\begin{pmatrix}
0&0&0&0&0\\
0&0&0&0&0\\
0&0&0&H_u&H_d\\
0&0&0&0&0\\
0&0&0&S&0
\end{pmatrix},\ 
\phi_2=\begin{pmatrix}
0&0&Q&0&0\\
0&0&L&0&0\\
0&0&0&0&0\\
0&0&0&0&0\\
0&0&0&0&0
\end{pmatrix},\ 
\phi_3=\begin{pmatrix}
0&0&0&0&0\\
0&0&0&0&0\\
0&0&0&0&0\\
U&N&0&0&0\\
D&E&0&0&0
\end{pmatrix}. 
\end{equation*}
Thanks to the structural interplay between chirality projections 
caused by the magnetic fluxes~(\ref{eq:magmag}) and the above orbifold 
projections~(\ref{eq:po}), almost all the phenomenologically unwanted 
extra massless modes are eliminated in the 4D spectrum, except for 
two generations of chiral exotic modes $S$ which  have the same 
hypercharge as the right-handed electrons. We also find that, 
in another similar pattern of suitable $T^6/Z_2\times Z'_2$ orbifold, 
these exotics remain after all. In the following, we propose a way 
to eliminate them for getting realistic models. 

It seems that the simplest way to eliminate the exotics 
is to twist the orbifold boundary conditions. 
The zero-mode structure on twisted orbifolds is different from 
that on untwisted ones \cite{Abe:2013bca}, and it is also known 
that this twisting is equivalent to turning on gauge field background 
with a vortex configuration~\cite{Buchmuller:2015eya}. 
Although we will skip the detail here, we explain only the essence 
we need for our purpose, that is, how to eliminate the exotics 
without spoiling the nonvanishing Yukawa couplings among quarks, 
leptons and Higgs bosons obtained above. 

Let us consider a twisted boundary condition of orbifolding in 
${\rm Re\,}z_3$ or ${\rm Im\,}z_3$ direction, only in the $U(1)_{r'}$ sector. 
The twisting phase is uniquely specified on $Z_2$ orbifolds. 
We notice that only the exotics $S$ feel a vanishing magnetic flux on 
the third 2D torus in the $U(1)_{r'}$ sector. As a consequence their 
zero-modes are eliminated by the twist, because the vanishing flux 
gives rise to a flat zero-mode wavefunction which cannot satisfy the 
twisted boundary condition. As for the other $U(1)_{r'}$ charged fields, 
$D$, $E$ and $H_d$, the number of their zero-modes are not changed 
since they feel $|M|=1$ unit of magnetic fluxes. All of the other 
$U(1)_{r'}$ singlet fields are obviously unaffected by this twisting. 
Thus, we can eliminate the exotic field $S$ in the low energy spectrum 
and derive a MSSM-like model without any of massless extra fields. 
This is one of the great feature of our model.

The VEVs of the right-handed sneutrinos lead to 
the lepton number violating term, 
\begin{equation}
y^\nu\langle\tilde\nu_R\rangle LH_u, 
\label{eq:mlhu}
\end{equation}
in the superpotential. 
This term clearly breaks the usual R-parity 
but we can find that our model has another discrete symmetry 
to prohibit the rapid proton decay. 
That is a $Z_3$ symmetry, so-called baryon triality~\cite{Ibanez:1991pr}, 
under which the MSSM fields transform in accordance with the charge 
assignment shown in Table~\ref{tb:z3p}. 
\begin{table}[t]
\center
\begin{tabular}{ccccccccc}
$Q$&$U$&$D$&$L$&$N$&$E$&$H_u$&$H_d$\\\hline
$1$&$\alpha^2$&$\alpha$&$\alpha^2$&$1$&$\alpha^2$&$\alpha$&$\alpha^2$\\
\end{tabular}
\caption{We show the transformation low of the MSSM fields 
under the $Z_3$ symmetry with $\alpha={\rm Exp}(2\pi i/3)$.}
\label{tb:z3p}
\end{table} 
This symmetry allows the presence of the $\mu$-term and lepton number violating terms 
but not baryon number violations, suppressing the proton decay. 
Within the MSSM matter contents, this can be an anomaly-free discrete 
gauge symmetry. 
Discrete symmetries without anomaly means that such symmetries can not be 
violated even by non-perturbative effects.
%This means that the $Z_3$ symmetry can be derived 
%as a remnant of a continuous anomaly-free $U(1)$ gauge symmetry. 
Although our model contains extra heavy Higgs fields other than 
the MSSM Higgs fields, this $Z_3$ symmetry can be anomaly-free 
because the copies of $(H_u, H_d)$ cannot contribute to the anomaly 
with the charge assignment shown in Table~\ref{tb:z3p}.\footnote{
See for anomalies of discrete symmetries~\cite{Araki:2008ek,Ishimori:2010au} and references therein.} 
We stop discussing the whole anomalies here because 
it is necessary to construct a full system which contains 
hidden sectors as well as the MSSM sector to study them completely.

\subsection{Mass eigenstates with sneutrino VEVs}
\label{ssec:snv}
We study the phenomenological impact of the lepton number violating mass 
term~(\ref{eq:mlhu}) in the superpotential. The total superpotential of our 
model is given by 
\begin{equation*}
W=y^{u}_{ijm}Q_iU_j{H_u}_m+y^{d}_{ijm}Q_iD_j{H_d}_m+
y^{\nu}_{ijm}L_iN_j{H_u}_m+y^{e}_{ijm}L_iE_j{H_d}_m+\mu_{mn}{H_u}_m{H_d}_n
+\tilde M_{im}L_i{H_u}_m, 
\end{equation*}
where $i,j=1,2,3$ and $m,n=1,2,3,4,5$. 
The right-handed neutrino superfield $N_j$ represents the fluctuation 
around the vacuum with nonvanishing sneutrino VEVs in the D-flat direction 
satisfying Eq.~(\ref{eq:modsusy}). 
The last term violating the lepton number is generated by 
the VEVs of sneutrinos $\tilde\nu_j$ as 
\begin{equation}
\tilde M_{im}L_i{H_u}_m=\left(y^{\nu}_{i1m}\langle\tilde\nu_1\rangle
+y^{\nu}_{i2m}\langle\tilde\nu_2\rangle
+y^{\nu}_{i3m}\langle\tilde\nu_3\rangle\right)
L_i{H_u}_m. 
\label{eq:lhmass}
\end{equation}
On the other hand, the second last term, so-called $\mu$-term, 
does not appear perturbatively in the 10D SYM theory compactified on tori, 
but is necessary for realizing EW symmetry breaking and some other 
phenomenological reasons. Because our magnetized SYM model is expected 
to be embedded into some D-brane configurations or other stringy setup, 
we assume that certain nonperturbative effects,\footnote{
See e.g. \cite{Blumenhagen:2006xt,Ibanez:2006da,Cvetic:2007ku,Kobayashi:2015siy}.} 
higher-dimensional 
operators or some other extrinsic effects generate this term 
in our model and here treat the components of this 
$(5\times 5)$-matrix $\tilde M_{im}$ as parameters. 

Let us consider the following rotation of the basis to diagonalize the 
mass terms of $H_u$,$H_d$ and $L$, 
\begin{eqnarray}
H'_u={\cal U}H_u,\qquad 
\begin{pmatrix}
L'\\H'_d
\end{pmatrix}
={\cal V}\begin{pmatrix}
L\\H_d
\end{pmatrix},
\label{eq:rotating}
\end{eqnarray}
where ${\cal U}$ and ${\cal V}$ are $(5\times5)$- and $(8\times8)$-unitary 
matrices, respectively. The superpotential~(\ref{eq:lhmass}) is rewritten as 
\begin{equation}
W=y'^{u}_{ijm}Q_iU_j{H'_u}_m+y'^{d}_{ijm}Q_iD_j{H'_d}_m
+y'^{\nu}_{ijm}L'_iU_j{H'_u}_m+y'^{e}_{ijm}L'_iE_j{H'_d}_m
+\tilde\mu_{mn}{H'_u}_m{H'_d}_n+W_{L\!\!\!/}, 
\label{eq:diagw}
\end{equation}
where 
\begin{equation}
\tilde\mu_{mn}={\rm diag}(m_1, m_2,\ldots,m_5). 
\label{eq:diagmu}
\end{equation}
We identify ${H'_u}_1$ and ${H'_d}_1$ with the MSSM Higgs fields, 
and the first entry $m_1$ in Eq.~(\ref{eq:diagmu}) corresponds to 
the $\mu$-parameter of the MSSM. 
The other ${H'_u}_{m \ne 1}$ and ${H'_d}_{m \ne 1}$ must be heavy 
enough to suppress the flavor changing neutral currents (FCNCs) 
and we assume $m_{m \ne 1}\gtrsim\mathcal O(10{\rm TeV})$. 

The last term $W_{L\!\!\!/}$ in the superpotential~(\ref{eq:diagw}) 
represents the lepton number violating terms 
in the present diagonal basis of $H'_u$,$H'_d$ and $L'$, 
\begin{equation}
W_{L\!\!\!/}
=\lambda_1QDL'+ \lambda_2L'L'E+\lambda_3NH'_uH'_d+\lambda_4EH'_dH'_d, 
\label{eq:wlp}
\end{equation}
those are generated from the Yukawa couplings in the original basis. 
Thanks to the baryon triality, we need not concern about the proton 
decay process caused by $W_{L\!\!\!/}$. However, studying the effect 
of these terms on the collider physics is interesting, because 
we can explicitly calculate all of the coupling constants except for 
the $\mu$-parameter in our model. This is one of the attractive features 
for building models based on magnetized toroidal compactifications. 

The Yukawa couplings 
$y^{u}_{ijm}$, $y^{d}_{ijm}$, $y^{\nu}_{ijm}$ and $y^{e}_{ijm}$ 
in the original basis are determined by 
the magnetic fluxes~(\ref{eq:magmag}) 
and the projection operators~(\ref{eq:po}), 
and their analytic forms can be derived. 
With these Yukawa couplings, the VEVs of multiple Higgs fields 
${H_u}_m$ and ${H_d}_m$ generate the mass matrices of 
quarks and leptons, e.g., 
\begin{equation*}
\left(M_u\right)_{ij}=y^{u}_{ij1}\langle {H_u}_1\rangle 
+y^{u}_{ij2}\langle {H_u}_2\rangle
+y^{u}_{ij3}\langle {H_u}_3\rangle
+y^{u}_{ij4}\langle {H_u}_4\rangle
+y^{u}_{ij5}\langle {H_u}_5\rangle, 
\end{equation*}
in the original basis. 
Here we remark that the sneutrino VEVs deform the lepton mass 
matrices through the diagonalization of $H_u$,$H_d$ and $L$, 
because the part of the diagonalizing matrix ${\cal V}$ which 
rotates $L_i$ in Eq.~(\ref{eq:rotating}) can be non-unitary 
(even though the whole $(8\times8)$-matrix ${\cal V}$ is unitary). 
We show this explicitly in the following based on a simplified situation. 
We here note that such an effect will be taken into account when 
we evaluate the mass ratios and mixing angles of quarks and leptons 
in Section~\ref{ssec:hv}. 

We can extract informations about the desired structure of the 
matrix $\mu_{mn}$ which we expect to be generated by extrinsic effects. 
For such a purpose, let us consider a simplified situation. 
We denote the relevant part of the superpotential by 
\begin{equation*}
W_{LH}=\mu_{mn}{H_u}_m{H_d}_n+\tilde M_{im}L_i{H_u}_m. 
\end{equation*}
Instead of the rotation~(\ref{eq:rotating}), 
let us suppose that the rotation of 
\begin{eqnarray}
{H_u}_m&\rightarrow&{H'_u}_m={\cal U}^{(u)}_{mn}{H_u}_n,\nonumber\\
{H_d}_m&\rightarrow&{H'_d}_m={\cal U}^{(d)}_{mn}{H_d}_n,\label{eq:lrot}\\
L_i&\rightarrow&L'_i={\cal V}_{ij}L_j,\nonumber
\end{eqnarray}
can diagonalize the matrices $\mu_{mn}$ and $\tilde M_{im}$ 
simultaneously as 
\begin{equation*}
W_{LH}= \sum_{i=1}^3\left(
\mu'_{i}{H'_u}_i{H'_d}_i
+\tilde M'_{i}L'_i{H'_u}_i\right)
+\sum_{q=4}^5\mu'_{q}{H'_u}_q{H'_d}_q, 
\end{equation*} 
where $\mu'_{i}$ and $\mu'_{q}$ are the eigenvalues of matrix $\mu_{mn}$. 
The three singular values of $( 3\times5)$-matrix $\tilde M_{im}$ 
are represented by $\tilde M'_{i}$, which we can calculate explicitly 
on concrete magnetized backgrounds. 

After the subsequent rotation, 
\begin{eqnarray*}
{H'_d}_i\rightarrow{H''_d}_i&\equiv&\frac1{\sqrt{(\mu'_i)^2+(\tilde M'_i)^2}}
\left(\mu'_i{H'_d}_i+\tilde M'_iL'_i\right),\\
L'_i\rightarrow L''_i&\equiv&\frac1{\sqrt{(\mu'_i)^2+(\tilde M'_i)^2}}
\left(-\tilde M'_i{H'_d}_i+\mu'_iL'_i\right),\\
\end{eqnarray*}
we find the final form of the superpotential as 
\begin{equation*}
W_{LH}= \sum_{i=1}^3\sqrt{(\mu'_i)^2+(\tilde M'_i)^2}{H'_u}_i{H''_d}_i
+\sum_{q=4}^5\mu'_{q}{H'_u}_q{H'_d}_q. 
\end{equation*} 
As we mentioned, in this diagonal basis of $L''$, $H'_u$ and $H''_d$, 
the mass matrices of charged leptons and neutrinos are deformed, e.g., 
\begin{equation*}
\left(M_e\right)_{ij}=
{\cal M}_{ik}{\cal V}^*_{kl}\left(y^{e}_{lj1}\langle {H_d}_1\rangle 
+y^{e}_{lj2}\langle {H_d}_2\rangle
+y^{e}_{lj3}\langle {H_d}_3\rangle
+y^{e}_{lj4}\langle {H_d}_4\rangle
+y^{e}_{lj5}\langle {H_d}_5\rangle\right), 
\end{equation*}
where the unitary matrix ${\cal V}_{ij}$ 
is given in Eq.~(\ref{eq:lrot}), and 
\begin{equation}
{\cal M}=\begin{pmatrix}
\frac{\mu'_1}{\sqrt{(\mu'_1)^2+(\tilde M'_1)^2}}&0&0\\
0&\frac{\mu'_2}{\sqrt{(\mu'_2)^2+(\tilde M'_2)^2}}&0\\
0&0&\frac{\mu'_3}{\sqrt{(\mu'_3)^2+(\tilde M'_3)^2}}\\
\end{pmatrix}.\label{eq:smat}
\end{equation}
Matrices ${\cal M}$ and ${\cal V}$ can change the mass eigenvalues 
and mixing angles of the leptons because their product ${\cal MV}$ 
is not unitary. 

In general, the largest one among the three values of $\tilde M'_i$ 
is of $\mathcal O(1)\times\langle\tilde\nu\rangle$, where 
$\langle\tilde\nu\rangle$ represents the typical scale of sneutrino VEVs, 
and a numerical analysis tells us that 
the other two values cannot be smaller than 
$\mathcal O(10^{-3})\times\langle\tilde\nu\rangle$ 
in a wide parameter space. 
The scale of $\langle\tilde\nu\rangle$ is comparable to the 
compactification scale because the flux-induced FI-terms are 
canceled by the sneutrino VEVs in our model. 
If $\langle\tilde\nu\rangle\sim M_{\rm GUT}\sim10^{16}$ GeV, 
the values of $\tilde M'_i$ are realized to be inside $10^{13}\sim10^{16}$ GeV. 
In this case, we want also $\mu'_i$ ($i=1,2,3$) to be so heavy because 
the effective Yukawa couplings of left-handed leptons $L''$ have the 
following factor
\begin{equation}
\mu'_i/\sqrt{(\mu'_i)^2+(\tilde M'_i)^2}. \label{eq:Lsup}
\end{equation}
When $\mu'_i\ll\tilde M'_i$, this would lead to an exceeding suppression, 
which causes some problems, clearly, in the charged-lepton sector. 

For this reason, we expect that the matrix $\mu_{mn}$ has the typical 
scale of ${\cal O}(M_{\rm GUT})$, and the rank of $\mu_{mn}$ is required 
to be at least three (the full rank is five). In order to suppress the 
FCNC due to the extra Higgs multiplets and realize a `natural' SUSY scenarios, 
one finds that the desirable rank of this matrix is four. 
We can then identify either $\left\{{H'_u}_4,{H'_d}_4\right\}$ or 
$\left\{{H'_u}_5,{H'_d}_5\right\}$ with the MSSM Higgs doublets, 
because the others must be heavy owing to $\tilde M'_i$ discussed above. 
When the matrix $\mu_{mn}$ is rank deficient, we can further infer its structure, 
because massive linear combinations of ${H_u}_m$ indicated by the matrix 
$\tilde M_{im}$ must also be mass eigenstates of $\mu_{mn}$ with nonvanishing 
mass eigenvalues. Otherwise, some or all of the left-handed leptons are 
decoupled from the other MSSM matter fields. 
This clearly restricts the texture of the matrix $\mu_{mn}$. 
Although we are studying the simplified situation given in Eq.~(\ref{eq:lrot}), 
a similar discussion could be available also in more general cases. 

We can adopt an alternative scenario with tiny values of the neutrino 
Yukawa couplings. It is known that a global suppression factor of 
Yukawa couplings can be induced in some special SYM systems compactified 
on magnetized tori, and the suppression can be strong enough to explain 
the tiny neutrino masses~\cite{Sumita:2015sta}. 
In the case with the tiny neutrino Yukawa couplings, 
the mass of $\tilde M_{im}L_i{H_u}_m$, is very light even when 
$\langle\tilde\nu\rangle\sim M_{\rm GUT}$, because the mass is 
given by a product of neutrino Yukawa couplings and sneutrino VEVs 
as shown in Eq.~(\ref{eq:lhmass}). 
In this case, the mass of $\mu_{mn}{H_u}_m{H_u}_n$ dominates $\tilde M'_i$, 
and there is no constraint on $\mu_{mn}$ to avoid exceeding suppressions 
of lepton Yukawa matrices as discussed below Eq.~(\ref{eq:Lsup}). 
As a result, this alternative scenario permits that the typical scale of 
$\mu_{mn}$ can be much lower than $M_{\rm GUT}$ (but that should be at least 
$\mathcal O(10)$ TeV in order to avoid the dangerous FCNC processes). 
This is very different from the previous scenario. 

For example, let us consider the case that a suppression factor of 
$\mathcal O(10^{-12})$ is realized for the neutrino Yukawa couplings. 
The heaviest neutrino (Dirac) mass is then estimated as 
$m_\nu\sim v_u\times\mathcal O(10^{-12})\sim O(0.1)$ eV 
($v_u$ is the VEV of the up-type Higgs field of the MSSM). 
We find the three singular values $\tilde M'_i$ are roughly of 
$\mathcal O(10\sim10^{4})$ GeV, which cannot induce the destructive 
suppression in the factor~(\ref{eq:Lsup}) even when $\mu'_i$ 
is comparable to the electroweak scale. 
For the case with $\mu_{mn}\gg\mathcal O(10)$ TeV, the natural 
SUSY scenarios require that the rank of $\mu_{mn}$ should be four. 
However, in the case with $\mu_{mn}\sim\mathcal O(10)$ TeV, 
the full rank of matrix $\mu_{mn}$ might be consistent with 
a low-scale SUSY breaking scenarios. 

\subsection{Mass ratios and mixing angles of quarks and leptons}
\label{ssec:hv}
We have constructed a MSSM-like model with the concrete configuration 
of magnetic fluxes~(\ref{eq:magmag}) on the $Z_2\times Z_2'$ orbifold 
characterized by the projection operators~(\ref{eq:po}). Finally, 
we study the masses and mixing angles of quarks and leptons in our model.  

The 4D effective Yukawa couplings in magnetized orbifold models 
can be expressed as linear combinations of $\eta_N$ defined by 
\begin{equation}
\eta_N\equiv \vartheta
\begin{bmatrix}
N/M\\0
\end{bmatrix}(0,\tau M),\label{eq:eta}
\end{equation}
where $M$ and $N$ are determined by the magnetic fluxes, 
and the Jacobi-theta function is given by 
\begin{equation*}
\vartheta
\begin{bmatrix}
a\\b
\end{bmatrix}(p,q)
=\sum_{\ell\in\mathbb{Z}}e^{\pi i(a+\ell)^2q}e^{2\pi i(a+\ell)(b+p)}. 
\end{equation*}
In our model, the parameter $\tau$ in Eq.~(\ref{eq:eta}) is identified 
with the complex structure of the first 2D torus in $(z_1,~\bar z_1)$ directions 
where the flavor structure of SM is produced. The parameter $M$ is given by 
a product of the effective magnetic fluxes felt by the left-handed matter, 
the right-handed matters and the Higgs fields on this 2D torus. Specifically, 
for the up-type quarks, the value of $M$ is $4\times8\times12=384$. 
It is similarly given as $M=180, 420$ and $180$ for the down-type quarks, 
neutrinos and charged leptons, respectively. With these numerical values, 
certain suitable hierarchies for the masses and the mixing angles can be 
reasonably obtained thanks to the Gaussian profile in 
$\eta_N$~\cite{Abe:2008sx, Abe:2014vza}. 

The analytic forms of the Yukawa couplings (before the right handed 
sneutrinos develop their VEVs) are explicitly shown in Appendix~\ref{sec:app}. 
These also determine the texture of the lepton number violating mass 
$\tilde M$ (\ref{eq:lhmass}) as discussed in the previous section. 
Based on them, we analyze the mass ratios of quarks and charged leptons 
as well as Cabibbo-Kobayashi-Maskawa (CKM)~\cite{Kobayashi:1973fv} and 
Pontecorvo-Maki-Nakagawa-Sakata (PMNS)~\cite{Pontecorvo:1967fh} 
mixing angles but not neutrino mass squared differences here, because 
the neutrino mass spectrum depends on whether they are Dirac or Majorana. 

The numerical analyses are simply performed for some sample values of 
parameters $\tau$, ${\cal M}_{ii}$ and VEVs $\langle\tilde\nu_i\rangle$, 
$v_{um}=\langle{H_u}_m\rangle$ and $v_{dm}=\langle{H_d}_m\rangle$. 
Note that, ${\cal M}_{ii}$ is a diagonal entry of matrix ${\cal M}$ 
given in Eq.~(\ref{eq:smat}), and it can be controlled by $\tilde \mu_i$ 
although the value of $\tilde M'_i$ is determined by the other parameters. 
The VEVs must satisfy the following conditions, 
\begin{equation*}
\sum_{i=1}^3\langle\tilde\nu_i\rangle^2=1,\qquad 
\sum_{m=1}^5 v_{um}^2=v_u^2,\qquad 
\sum_{m=1}^5 v_{dm}^2=v_d^2, 
\end{equation*}
where $v_u$ and $v_d$ are the VEVs of the MSSM Higgs doublets. 
The first one is required to satisfy the modified SUSY 
condition~(\ref{eq:modsusy}) on the magnetized background~(\ref{eq:magmag}). 
Since our interest here is the mass ratios and mixing angles, 
the ratios of these VEVs are important. 
Note also that, the effects of renormalization group equations (RGEs) 
are not included in the analysis. 

\begin{table}[t]
\begin{center}
\begin{tabular}{|c||c|c|} \hline
 & Sample values & Observed \\ \hline
$(m_u/m_t, m_c/m_t)$ & 
$(6.6 \times 10^{-5}, 6.8\times10^{-2})$ & 
$(1.3 \times 10^{-5}, 7.4\times10^{-3})$ 
\\ \hline
$(m_d/m_b, m_s/m_b )$ & 
$(2.0 \times 10^{-4}, 3.5\times10^{-2})$ & 
$(1.1 \times 10^{-3}, 2.3\times10^{-2})$ 
\\ \hline
$(m_e/m_\tau, m_\mu/m_\tau )$ & 
$(1.7 \times 10^{-3}, 1.6\times 10^{-2})$ & 
$(2.9 \times 10^{-4}, 6.0\times10^{-2})$ 
\\ \hline \hline 
$|V_{\rm CKM}|$ & 
\begin{minipage}{0.35\linewidth}
\begin{eqnarray} 
\left( 
\begin{array}{ccc}
0.97 & 0.24 & 0.030 \\
0.23 & 0.95 & 0.22 \\
0.081 & 0.20 & 0.98 
\end{array}
\right) 
\nonumber
\end{eqnarray} \\*[-20pt]
\end{minipage}
& 
\begin{minipage}{0.4\linewidth}
\begin{eqnarray} 
\left( 
\begin{array}{ccc}
0.97 & 0.23 & 0.0035 \\
0.23 & 0.97 & 0.041 \\
0.0087 & 0.040 & 1.0 
\end{array}
\right) 
\nonumber
\end{eqnarray} \\*[-20pt]
\end{minipage} 
\\ \hline \hline
$|V_{\rm PMNS}|$ & 
\begin{minipage}{0.35\linewidth}
\begin{eqnarray} 
\left( 
\begin{array}{ccc}
0.91 & 0.37 & 0.13 \\
0.32 & 0.90 & 0.29 \\
0.23 & 0.23 & 0.95 
\end{array}
\right) 
\nonumber
\end{eqnarray} \\*[-20pt]
\end{minipage}
& 
\begin{minipage}{0.4\linewidth}
\begin{eqnarray} 
\left( 
\begin{array}{ccc}
0.82 & 0.55 & 0.16 \\
0.51 & 0.58 & 0.64 \\
0.26 & 0.61 & 0.75 
\end{array}
\right) 
\nonumber
\end{eqnarray} \\*[-20pt]
\end{minipage} \\ \hline
\end{tabular}
\end{center}
\caption{The sample theoretical values of the mass ratios of 
quarks and charged leptons as well as CKM and PMNS mixing 
matrices are shown. We quote the observed values from 
Ref.~\cite{Beringer:1900zz}.}
\label{tab:massmix}
\end{table}

It is found that the following set 
\begin{eqnarray*}
\tau&=&5i,\qquad \langle\tilde\nu_3\rangle=1,\qquad 
{\cal M}_{22}/{\cal M}_{11}=3, \qquad {\cal M}_{33}/{\cal M}_{11}=25,\\ 
v_{u3}/v_{u5}&=&2,\qquad v_{u4}/v_{u5}=4,\qquad 
v_{d3}/v_{d2}=12,\qquad v_{d4}/v_{d2}=14, 
\end{eqnarray*}
and 
$\langle\tilde\nu_1\rangle
=\langle\tilde\nu_2\rangle=v_{u0}=v_{u1}=v_{d0}=v_{d4}=0$ 
leads to a hopeful pattern of the mass ratios and the mixing angles 
as shown in Table \ref{tab:massmix}. 
Although there are some unacceptable deviations from the 
observed values especially in the quark sector, we remark 
that these theoretical values are derived from very limited 
sample choices of parameters. We expect that  more realistic 
pattern would be obtained by thorough analyses which remain 
as future works. 
It is interesting that the observed Cabibbo angle is obtained 
even in this simple analysis, that is almost unchanged by 
the RGE effects~\cite{Sasaki:1986jv}.

\section{Summary}
\label{sec:summary}
We have studied a new class of supersymmetric models on magnetized 
orbifold, where the nonvanishing FI-terms are induced by magnetic 
fluxes in the extra compact space. Scalar fields charged under the 
fluxed gauge symmetries tend to develop nonvanishing VEVs in the D-flat 
directions to cancel the FI-terms and SUSY is recovered on such vacua. 
This idea has broadened the variety of magnetized models. Especially, 
as a concrete phenomenological example, we have analyzed the case that 
the fluxed gauge symmetries possessing nonvanishing FI-terms are two 
$U(1)$ symmetries under which the right-handed neutrinos are charged. 
In this case, the sneutrino VEVs along the D-flat directions cancel 
the FI-terms out leading to a new class of SUSY vacua, where 
all the unwanted chiral exotics and massless adjoint fields, 
which generically appear in string or string-inspired models, 
are eliminated completely. 

We have also studied the phenomenology of this model focusing on the 
effects of right-handed sneutrino VEVs which induce a mass term 
$LH_u$ in the superpotential. It violates the lepton number, 
and our model does not have the R-parity that is usually assumed to 
suppress the proton decay. Instead, we have found that our model has 
the $Z_3$ symmetry called the baryon triality, which forbids baryon 
number violating processes and ensures the long life-time of proton 
to be consistent with the non-observations of its decay. 
After diagonalizing the whole mass matrices of $L$, $H_u$ and $H_d$, 
the lepton number violating masses are eliminated and consequently 
the lepton flavor structure is modified in this new basis. 
Such a correction for leptonic Yukawa couplings is determined by 
the interplay between the lepton number violating mass and 
the SUSY Higgs mass (so called $\mu$-parameters) in the superpotential. 

By introducing Higgs VEVs, we have finally performed a rough analysis 
for parameters those yield semi-realistic flavor structures, 
and shown sample theoretical values of mass ratios and mixing angles 
of quarks and leptons. Yukawa couplings in certain magnetized orbifold 
models have a texture which induces suitable hierarchies reasonably 
and yields a semi-realistic pattern of the hierarchies without 
hierarchical input parameters~\cite{Abe:2014vza}. 
The texture is modified in our model due to the lepton number 
violating mass. Although the rough estimation in Section~\ref{ssec:hv} 
shows some deviations from the observed values, we expect that 
they would be improved by the thorough analyses in future works, 
where CP-violating phases should also be studied 
(CP-violating phases of the quark sector 
in the magnetized orbifold models were recently studied 
in Ref.~\cite{Kobayashi:2016qag}).

Accepting flux-induced FI-terms provides a new class of SUSY models 
in SYM theories compactified on magnetized tori/orbifolds. 
We expect that the scenario of cancellation between the FI-terms and 
the VEVs of bifundamental fields can be applied to the other model 
building for visible (e.g., from other gauge groups~\cite{Choi:2009pv}), 
hidden (e.g., SUSY breaking~\cite{Abe:2016zgq}) 
and moduli stabilization sectors. Furthermore, if the localized 
fluxes like vortex configurations~\cite{Buchmuller:2015eya} also 
contribute to the FI-terms, it might be possible to generate 
nontrivial wavefunction profiles of charged fields~\cite{Lee:2003mc} 
as in the five-dimensional SUSY~\cite{GrootNibbelink:2002wv} 
and supergravity~\cite{Abe:2004yk} models. It is also 
interesting to consider the fluxed $U(1)$ symmetry to be anomalous, 
and study the combination of flux-induced FI-terms and loop-corrected 
ones~\cite{Green:1984sg} caused by the anomaly. 

\subsection*{Acknowledgement}
Y.~T. would like to thank Yusuke Shimizu for useful comments. 
H.~A. was supported by JSPS KAKENHI Grant Number JP16K05330. 
T.~K. was supported in part by JSPS KAKENHI Grant Number JP26247042. 
K.~S. was supported by Waseda University Grant 
for Special Research Projects No.~2016B-200. 
Y.~T. was supported by JSPS KAKENHI Grant Number JP16J04612.

%\section*{Appendix}
%\addcontentsline{toc}{section}{Appendix}
%\appendix
%\section{hoge}

\appendix
\section{Yukawa couplings}
\label{sec:app}

We show the analytic forms of Yukawa couplings for the model 
given in Section~\ref{ssec:3gen}, where all the (super)fields 
$Q_i$, $U_j$, $D_i$, $L_i$, $N_j$, $E_j$, ${H_u}_m$ and ${H_d}_m$ 
are in their original basis before the sneutrinos develop VEVs. 

For the quark sector, 
the Yukawa couplings involving $Q_i$, $U_j$ and ${H_u}_m$ are given by 
\begin{eqnarray}
y^u_{ij1}&=&\begin{pmatrix}
y_b&0&-y_l\\
0&\frac1{\sqrt2}(y_e-y_i)&0\\
-y_f&0&y_h
\end{pmatrix},\qquad
y^u_{ij2}=\begin{pmatrix}
0&y_c-y_k&0\\
\frac1{\sqrt2}(y_b-y_h)&0&\frac1{\sqrt2}(y_f-y_l)\\
0&0&0
\end{pmatrix},\nonumber\\
y^u_{ij3}&=&\begin{pmatrix}
-y_j&0&y_d\\
0&\frac1{\sqrt2}(y_a-y_m)&0\\
y_d&0&-y_j
\end{pmatrix},\qquad
y^u_{ij4}=\begin{pmatrix}
0&0&0\\
\frac1{\sqrt2}(y_f-y_l)&0&\frac1{\sqrt2}(y_b-y_h)\\
0&y_c-y_k&0
\end{pmatrix},\nonumber\\
y^u_{ij5}&=&\begin{pmatrix}
y_h&0&-y_f\\
0&\frac1{\sqrt2}(y_e-y_i)&0\\
-y_l&0&y_b
\end{pmatrix},
\label{eq:yu}
\end{eqnarray}
where 
\begin{eqnarray}
y_a&=&\eta_0+\eta_{96}+\eta_{192}+\eta_{96},\qquad
y_b=\eta_4+\eta_{100}+\eta_{188}+\eta_{92},\nonumber\\
y_c&=&\eta_8+\eta_{104}+\eta_{184}+\eta_{88},\qquad
y_d=\eta_{12}+\eta_{108}+\eta_{180}+\eta_{84},\nonumber\\
y_e&=&\eta_{16}+\eta_{112}+\eta_{176}+\eta_{80},\qquad
y_f=\eta_{20}+\eta_{116}+\eta_{172}+\eta_{76},\nonumber\\
y_g&=&\eta_{24}+\eta_{120}+\eta_{168}+\eta_{72},\qquad
y_h=\eta_{28}+\eta_{124}+\eta_{164}+\eta_{68},\nonumber\\
y_i&=&\eta_{32}+\eta_{128}+\eta_{160}+\eta_{64},\qquad
y_j=\eta_{36}+\eta_{132}+\eta_{156}+\eta_{60},\nonumber\\
y_k&=&\eta_{40}+\eta_{136}+\eta_{152}+\eta_{56},\qquad
y_l=\eta_{44}+\eta_{140}+\eta_{148}+\eta_{52},\nonumber\\
y_m&=&\eta_{48}+\eta_{144}+\eta_{144}+\eta_{48},\nonumber
\end{eqnarray}
while those among $Q_i$, $D_j$ and ${H_d}_m$ are given by 
\begin{eqnarray}
y^d_{ij1}&=&\begin{pmatrix}
\eta_{0}&\sqrt2\eta_{36}&\sqrt2\eta_{72}\\
\sqrt2\eta_{45}&\eta_{9}+\eta_{81}&\eta_{27}+\eta_{63}\\
\eta_{90}&\sqrt2\eta_{54}&\sqrt2\eta_{18}
\end{pmatrix},\nonumber\\
y^d_{ij2}&=&\begin{pmatrix}
\frac1{\sqrt2}(\eta_{20}+\eta_{40})&\eta_{4}+\eta_{76}&\eta_{32}+\eta_{68}\\
\eta_{5}+\eta_{85}&\frac1{\sqrt2}(\eta_{31}+\eta_{41}+\eta_{49}+\eta_{59})
&\frac1{\sqrt2}(\eta_{13}+\eta_{23}+\eta_{67}+\eta_{77})\\
\sqrt2\eta_{50}&\eta_{44}+\eta_{64}&\eta_{22}+\eta_{58}
\end{pmatrix},\nonumber\\
y^d_{ij3}&=&\begin{pmatrix}
\frac1{\sqrt2}(\eta_{20}+\eta_{40})&\eta_{44}+\eta_{64}&\eta_{8}+\eta_{28}\\
\eta_{35}+\eta_{55}&\frac1{\sqrt2}(\eta_{1}+\eta_{19}+\eta_{71}+\eta_{89})
&\frac1{\sqrt2}(\eta_{17}+\eta_{37}+\eta_{53}+\eta_{73})\\
\sqrt2\eta_{10}&\eta_{26}+\eta_{46}&\eta_{62}+\eta_{82}
\end{pmatrix},\nonumber\\
y^d_{ij4}&=&\begin{pmatrix}
\frac1{\sqrt2}(\eta_{60}+\eta_{80})&\eta_{24}+\eta_{84}&\eta_{12}+\eta_{48}\\
\eta_{15}+\eta_{75}&\frac1{\sqrt2}(\eta_{21}+\eta_{39}+\eta_{51}+\eta_{69})
&\frac1{\sqrt2}(\eta_{3}+\eta_{33}+\eta_{57}+\eta_{87})\\
\sqrt2\eta_{30}&\eta_{6}+\eta_{26}&\eta_{42}+\eta_{78}
\end{pmatrix},\nonumber\\
y^d_{ij5}&=&\begin{pmatrix}
\frac1{\sqrt2}(\eta_{60}+\eta_{80})&\eta_{16}+\eta_{56}&\eta_{52}+\eta_{88}\\
\eta_{25}+\eta_{65}&\frac1{\sqrt2}(\eta_{11}+\eta_{29}+\eta_{61}+\eta_{79})
&\frac1{\sqrt2}(\eta_{7}+\eta_{43}+\eta_{47}+\eta_{83})\\
\sqrt2\eta_{70}&\eta_{34}+\eta_{74}&\eta_{2}+\eta_{38}
\end{pmatrix}. 
\label{eq:yd}
\end{eqnarray}

For the lepton sector, the Yukawa couplings 
between $L_i$, $N_j$ and ${H_u}_m$ are given by 
\begin{eqnarray}
y^\nu_{ij1}&=&\frac1{\sqrt2}\begin{pmatrix}
\sqrt2(\eta_{5}-\eta_{65})&\sqrt2(\eta_{185}-\eta_{115})&
\sqrt2(\eta_{55}+\eta_{125})\\
\eta_{173}-\eta_{103}-\eta_{187}+\eta_{163}&
\eta_{67}-\eta_{137}-\eta_{53}+\eta_{17}&
\eta_{113}-\eta_{43}-\eta_{127}+\eta_{197}\\
\eta_{79}-\eta_{149}-\eta_{19}+\eta_{89}&
\eta_{101}-\eta_{31}-\eta_{199}+\eta_{151}&
\eta_{139}-\eta_{209}-\eta_{41}+\eta_{29}
\end{pmatrix},\nonumber\\
y^\nu_{ij2}&=&\frac1{\sqrt2}\begin{pmatrix}
\sqrt2(\eta_{170}-\eta_{110})&\sqrt2(\eta_{10}-\eta_{130})&
\sqrt2(\eta_{190}+\eta_{50})\\
\eta_{2}-\eta_{142}-\eta_{58}+\eta_{82}&
\eta_{178}-\eta_{38}-\eta_{122}+\eta_{158}&
\eta_{62}-\eta_{202}-\eta_{118}+\eta_{22}\\
\eta_{166}-\eta_{26}-\eta_{194}+\eta_{94}&
\eta_{74}-\eta_{206}-\eta_{46}+\eta_{94}&
\eta_{106}-\eta_{34}-\eta_{134}+\eta_{146}
\end{pmatrix},\nonumber\\
y^\nu_{ij3}&=&\frac1{\sqrt2}\begin{pmatrix}
\sqrt2(\eta_{75}-\eta_{135})&\sqrt2(\eta_{165}-\eta_{45})&
\sqrt2(\eta_{15}-\eta_{195})\\
\eta_{177}-\eta_{33}-\eta_{117}+\eta_{93}&
\eta_{3}-\eta_{207}-\eta_{123}+\eta_{87}&
\eta_{183}-\eta_{27}-\eta_{57}+\eta_{153}\\
\eta_{9}-\eta_{201}-\eta_{51}+\eta_{81}&
\eta_{171}-\eta_{39}-\eta_{129}+\eta_{81}&
\eta_{69}-\eta_{141}-\eta_{111}+\eta_{99}
\end{pmatrix},\nonumber\\
y^\nu_{ij4}&=&\frac1{\sqrt2}\begin{pmatrix}
\sqrt2(\eta_{100}-\eta_{140})&\sqrt2(\eta_{80}-\eta_{200})&
\sqrt2(\eta_{160}-\eta_{20})\\
\eta_{68}-\eta_{208}-\eta_{128}+\eta_{152}&
\eta_{172}-\eta_{32}-\eta_{52}+\eta_{88}&
\eta_{8}-\eta_{148}-\eta_{188}+\eta_{92}\\
\eta_{184}-\eta_{44}-\eta_{124}+\eta_{164}&
\eta_{4}-\eta_{136}-\eta_{116}+\eta_{164}&
\eta_{176}-\eta_{104}-\eta_{64}+\eta_{76}
\end{pmatrix},\nonumber\\
y^\nu_{ij5}&=&\frac1{\sqrt2}\begin{pmatrix}
\sqrt2(\eta_{145}-\eta_{205})&\sqrt2(\eta_{95}-\eta_{25})&
\sqrt2(\eta_{85}-\eta_{155})\\
\eta_{107}-\eta_{37}-\eta_{47}+\eta_{23}&
\eta_{73}-\eta_{143}-\eta_{193}+\eta_{157}&
\eta_{167}-\eta_{97}-\eta_{13}+\eta_{83}\\
\eta_{61}-\eta_{131}-\eta_{121}+\eta_{11}&
\eta_{179}-\eta_{109}-\eta_{59}+\eta_{11}&
\eta_{1}-\eta_{71}-\eta_{181}+\eta_{169}
\end{pmatrix}. 
\label{eq:yn}
\end{eqnarray}
As for the Yukawa couplings $y^e_{ijm}$ 
involving $L_i$, $E_j$ and ${H_d}_m$, 
the $3 \times 3$ matrices $y^e_{ijm}$ for each $m$ 
are equivalent to the corresponding transposed 
matrices for the down-type quarks, that is, 
$y^e_{ijm}=y^d_{jim}$.


\begin{thebibliography}{99}

%\cite{Aad:2012tfa}
\bibitem{Aad:2012tfa}
  G.~Aad {\it et al.} [ATLAS Collaboration],
  %``Observation of a new particle in the search for the Standard Model Higgs boson with the ATLAS detector at the LHC,''
  Phys.\ Lett.\ B {\bf 716} (2012) 1
 % doi:10.1016/j.physletb.2012.08.020
  [arXiv:1207.7214 [hep-ex]];
  %%CITATION = doi:10.1016/j.physletb.2012.08.020;%%
  %6405 citations counted in INSPIRE as of 05 Sep 2016
%
%\cite{Chatrchyan:2012xdj}
%\bibitem{Chatrchyan:2012xdj}
  S.~Chatrchyan {\it et al.} [CMS Collaboration],
  %``Observation of a new boson at a mass of 125 GeV with the CMS experiment at the LHC,''
  Phys.\ Lett.\ B {\bf 716} (2012) 30
 % doi:10.1016/j.physletb.2012.08.021
  [arXiv:1207.7235 [hep-ex]].
  %%CITATION = doi:10.1016/j.physletb.2012.08.021;%%
  %6258 citations counted in INSPIRE as of 05 Sep 2016

%\cite{ArkaniHamed:1999dc}
\bibitem{ArkaniHamed:1999dc}
  N.~Arkani-Hamed and M.~Schmaltz,
  %``Hierarchies without symmetries from extra dimensions,''
  Phys.\ Rev.\ D {\bf 61} (2000) 033005
 % doi:10.1103/PhysRevD.61.033005
  [hep-ph/9903417].
  %%CITATION = doi:10.1103/PhysRevD.61.033005;%%
  %658 citations counted in INSPIRE as of 23 Oct 2016

%\cite{Bachas:1995ik,Cremades:2004wa}
\bibitem{Bachas:1995ik}
  C.~Bachas,
  %``A Way to break supersymmetry,''
  hep-th/9503030.
  %%CITATION = HEP-TH/9503030;%%
  %329 citations counted in INSPIRE as of 19 Sep 2014

%\cite{Cremades:2004wa}
\bibitem{Cremades:2004wa}
  D.~Cremades, L.~E.~Ibanez and F.~Marchesano,
  %``Computing Yukawa couplings from magnetized extra dimensions,''
  JHEP {\bf 0405} (2004) 079  [hep-th/0404229].
  %%CITATION = HEP-TH/0404229;%%

%\cite{Abe:2012fj}
\bibitem{Abe:2012fj}
  H.~Abe, T.~Kobayashi, H.~Ohki, A.~Oikawa and K.~Sumita,
  %``Phenomenological aspects of 10D SYM theory with magnetized extra dimensions,''
  Nucl.\ Phys.\ B {\bf 870} (2013) 30
  %doi:10.1016/j.nuclphysb.2013.01.014
  [arXiv:1211.4317 [hep-ph]]. 
  %%CITATION = doi:10.1016/j.nuclphysb.2013.01.014;%%
  %20 citations counted in INSPIRE as of 19 Apr 2016

%\cite{Abe:2014soa}
\bibitem{Abe:2014soa}
  H.~Abe, J.~Kawamura and K.~Sumita,
  %``The Higgs boson mass and SUSY spectra in 10D SYM theory with magnetized extra dimensions,''
  Nucl.\ Phys.\ B {\bf 888} (2014) 194
  %doi:10.1016/j.nuclphysb.2014.09.016
  [arXiv:1405.3754 [hep-ph]].
  %%CITATION = doi:10.1016/j.nuclphysb.2014.09.016;%%
  %3 citations counted in INSPIRE as of 19 Apr 2016

%\cite{Abe:2009vi}
\bibitem{Abe:2009vi}
  H.~Abe, K.~S.~Choi, T.~Kobayashi and H.~Ohki,
  %``Non-Abelian Discrete Flavor Symmetries from Magnetized/Intersecting Brane Models,''
  Nucl.\ Phys.\ B {\bf 820} (2009) 317
  %doi:10.1016/j.nuclphysb.2009.05.024
  [arXiv:0904.2631 [hep-ph]];
  %%CITATION = doi:10.1016/j.nuclphysb.2009.05.024;%%
  %56 citations counted in INSPIRE as of 05 Sep 2016
%
%\cite{Abe:2009uz}
%\bibitem{Abe:2009uz}
  %H.~Abe, K.~S.~Choi, T.~Kobayashi and H.~Ohki,
  %``Magnetic flux, Wilson line and orbifold,''
  Phys.\ Rev.\ D {\bf 80} (2009) 126006
  %doi:10.1103/PhysRevD.80.126006
  [arXiv:0907.5274 [hep-th]];
  %%CITATION = doi:10.1103/PhysRevD.80.126006;%%
  %33 citations counted in INSPIRE as of 23 Oct 2016
%
%\cite{Abe:2010ii}
%\bibitem{Abe:2010ii}
  %H.~Abe, K.~S.~Choi, T.~Kobayashi and H.~Ohki,
  %``Flavor structure from magnetic fluxes and non-Abelian Wilson lines,''
  Phys.\ Rev.\ D {\bf 81} (2010) 126003
  %doi:10.1103/PhysRevD.81.126003
  [arXiv:1001.1788 [hep-th]];
  %%CITATION = doi:10.1103/PhysRevD.81.126003;%%
  %27 citations counted in INSPIRE as of 23 Oct 2016
%
%\cite{Abe:2014nla}
%\bibitem{Abe:2014nla}
  H.~Abe, T.~Kobayashi, H.~Ohki, K.~Sumita and Y.~Tatsuta,
  %``Non-Abelian discrete flavor symmetries of 10D SYM theory with magnetized extra dimensions,''
  JHEP {\bf 1406} (2014) 017
  %doi:10.1007/JHEP06(2014)017
  [arXiv:1404.0137 [hep-th]].
  %%CITATION = doi:10.1007/JHEP06(2014)017;%%
  %8 citations counted in INSPIRE as of 23 Oct 2016

%\cite{Abe:2008fi}
\bibitem{Abe:2008fi}
  H.~Abe, T.~Kobayashi and H.~Ohki,
  %``Magnetized orbifold models,''
  JHEP {\bf 0809} (2008) 043
  %doi:10.1088/1126-6708/2008/09/043
  [arXiv:0806.4748 [hep-th]].
  %%CITATION = doi:10.1088/1126-6708/2008/09/043;%%
  %39 citations counted in INSPIRE as of 16 Mar 2016

%\cite{Abe:2008sx}
\bibitem{Abe:2008sx}
  H.~Abe, K.~S.~Choi, T.~Kobayashi and H.~Ohki,
  %``Three generation magnetized orbifold models,''
  Nucl.\ Phys.\ B {\bf 814} (2009) 265
  %doi:10.1016/j.nuclphysb.2009.02.002
  [arXiv:0812.3534 [hep-th]].
  %%CITATION = doi:10.1016/j.nuclphysb.2009.02.002;%%
  %33 citations counted in INSPIRE as of 04 avril 2016

%\cite{Abe:2014noa}
\bibitem{Abe:2014noa}
  T.~h.~Abe, Y.~Fujimoto, T.~Kobayashi, T.~Miura, K.~Nishiwaki and M.~Sakamoto,
  %``Operator analysis of physical states on magnetized $T^{2}/Z_{N}$ orbifolds,''
  Nucl.\ Phys.\ B {\bf 890} (2014) 442
  %doi:10.1016/j.nuclphysb.2014.11.022
  [arXiv:1409.5421 [hep-th]];
  %%CITATION = doi:10.1016/j.nuclphysb.2014.11.022;%%
  %8 citations counted in INSPIRE as of 05 Sep 2016
%
%\cite{Abe:2015yva}
%\bibitem{Abe:2015yva}
  T.~h.~Abe, Y.~Fujimoto, T.~Kobayashi, T.~Miura, K.~Nishiwaki, M.~Sakamoto and Y.~Tatsuta,
  %``Classification of three-generation models on magnetized orbifolds,''
  Nucl.\ Phys.\ B {\bf 894} (2015) 374
  %doi:10.1016/j.nuclphysb.2015.03.004
  [arXiv:1501.02787 [hep-ph]];
  %%CITATION = doi:10.1016/j.nuclphysb.2015.03.004;%%
  %10 citations counted in INSPIRE as of 05 Sep 2016
%
%\cite{Matsumoto:2016okl}
%\bibitem{Matsumoto:2016okl}
  Y.~Matsumoto and Y.~Sakamura,
  %``Yukawa couplings in 6D gauge Higgs unification on T$^2$/Z$_N$ with magnetic fluxes,''
  PTEP {\bf 2016} (2016) no.5,  053B06
  %doi:10.1093/ptep/ptw058
  [arXiv:1602.01994 [hep-ph]];
  %%CITATION = doi:10.1093/ptep/ptw058;%%
  %2 citations counted in INSPIRE as of 05 Sep 2016
%
%\cite{Fujimoto:2016zjs}
%\bibitem{Fujimoto:2016zjs}
  Y.~Fujimoto, T.~Kobayashi, K.~Nishiwaki, M.~Sakamoto and Y.~Tatsuta,
  %``Comprehensive Analysis of Yukawa Hierarchies on $T^2/Z_N$ with Magnetic Fluxes,''
  Phys.\ Rev.\ D {\bf 94} (2016) no.3,  035031
  %doi:10.1103/PhysRevD.94.035031
  [arXiv:1605.00140 [hep-ph]].
  %%CITATION = doi:10.1103/PhysRevD.94.035031;%%
  %2 citations counted in INSPIRE as of 05 Sep 2016

%\cite{Fayet:1974jb}
\bibitem{Fayet:1974jb}
  P.~Fayet and J.~Iliopoulos,
  %``Spontaneously Broken Supergauge Symmetries and Goldstone Spinors,''
  Phys.\ Lett.\ B {\bf 51} (1974) 461.
  %doi:10.1016/0370-2693(74)90310-4
  %%CITATION = doi:10.1016/0370-2693(74)90310-4;%%
  %769 citations counted in INSPIRE as of 05 Sep 2016

%\cite{Abe:2013bba}
\bibitem{Abe:2013bba}
  H.~Abe, T.~Kobayashi, H.~Ohki, K.~Sumita and Y.~Tatsuta,
  %``Flavor landscape of 10D SYM theory with magnetized extra dimensions,''
  JHEP {\bf 1404} (2014) 007
  %doi:10.1007/JHEP04(2014)007
  [arXiv:1307.1831 [hep-th]].
  %%CITATION = doi:10.1007/JHEP04(2014)007;%%
  %8 citations counted in INSPIRE as of 19 Apr 2016

%\cite{Green:1984sg}
\bibitem{Green:1984sg}
  M.~B.~Green and J.~H.~Schwarz,
  %``Anomaly Cancellation in Supersymmetric D=10 Gauge Theory and Superstring Theory,''
  Phys.\ Lett.\  {\bf 149B} (1984) 117.
 % doi:10.1016/0370-2693(84)91565-X
  %%CITATION = doi:10.1016/0370-2693(84)91565-X;%%
  %2469 citations counted in INSPIRE as of 23 Oct 2016

%\cite{Marcus:1983wb,ArkaniHamed:2001tb}
\bibitem{Marcus:1983wb}
  N.~Marcus, A.~Sagnotti and W.~Siegel,
  %``Ten-dimensional Supersymmetric {Yang-Mills} Theory in Terms of Four-dimensional Superfields,''
  Nucl.\ Phys.\ B {\bf 224} (1983) 159.
  %%CITATION = NUPHA,B224,159;%%
  %95 citations counted in INSPIRE as of 02 Oct 2014

%\cite{ArkaniHamed:2001tb}
\bibitem{ArkaniHamed:2001tb}
  N.~Arkani-Hamed, T.~Gregoire and J.~G.~Wacker,
  %``Higher dimensional supersymmetry in 4-D superspace,''
  JHEP {\bf 0203} (2002) 055
  [hep-th/0101233].
  %%CITATION = HEP-TH/0101233;%%
  %250 citations counted in INSPIRE as of 02 Oct 2014

%\cite{Abe:2012ya}
\bibitem{Abe:2012ya}
  H.~Abe, T.~Kobayashi, H.~Ohki and K.~Sumita,
  %``Superfield description of 10D SYM theory with magnetized extra dimensions,''
  Nucl.\ Phys.\ B {\bf 863} (2012) 1
  %doi:10.1016/j.nuclphysb.2012.05.012
  [arXiv:1204.5327 [hep-th]].
  %%CITATION = doi:10.1016/j.nuclphysb.2012.05.012;%%
  %19 citations counted in INSPIRE as of 19 Apr 2016

%\cite{Abe:2014vza}
\bibitem{Abe:2014vza}
  H.~Abe, T.~Kobayashi, K.~Sumita and Y.~Tatsuta,
  %``Gaussian Froggatt-Nielsen mechanism on magnetized orbifolds,''
  Phys.\ Rev.\ D {\bf 90} (2014) no.10,  105006
  %doi:10.1103/PhysRevD.90.105006
  [arXiv:1405.5012 [hep-ph]].
  %%CITATION = doi:10.1103/PhysRevD.90.105006;%%
  %7 citations counted in INSPIRE as of 04 avril 2016

%\cite{Abe:2013bca}
\bibitem{Abe:2013bca}
  T.~H.~Abe, Y.~Fujimoto, T.~Kobayashi, T.~Miura, K.~Nishiwaki and M.~Sakamoto,
  %``$Z_N$ twisted orbifold models with magnetic flux,''
  JHEP {\bf 1401} (2014) 065
  %doi:10.1007/JHEP01(2014)065
  [arXiv:1309.4925 [hep-th]].
  %%CITATION = doi:10.1007/JHEP01(2014)065;%%
  %16 citations counted in INSPIRE as of 28 Jun 2016

%\cite{Buchmuller:2015eya}
\bibitem{Buchmuller:2015eya}
  W.~Buchmuller, M.~Dierigl, F.~Ruehle and J.~Schweizer,
  %``Chiral fermions and anomaly cancellation on orbifolds with Wilson lines and flux,''
  Phys.\ Rev.\ D {\bf 92} (2015) no.10,  105031
  %doi:10.1103/PhysRevD.92.105031
  [arXiv:1506.05771 [hep-th]].
  %%CITATION = doi:10.1103/PhysRevD.92.105031;%%
  %3 citations counted in INSPIRE as of 28 Jun 2016

%\cite{Ibanez:1991pr}
\bibitem{Ibanez:1991pr}
  L.~E.~Ibanez and G.~G.~Ross,
  %``Discrete gauge symmetries and the origin of baryon and lepton number conservation in supersymmetric versions of the standard model,''
  Nucl.\ Phys.\ B {\bf 368} (1992) 3.
  %doi:10.1016/0550-3213(92)90195-H
  %%CITATION = doi:10.1016/0550-3213(92)90195-H;%%
  %521 citations counted in INSPIRE as of 06 Sep 2016
  
%\cite{Araki:2008ek}
\bibitem{Araki:2008ek} 
  T.~Araki, T.~Kobayashi, J.~Kubo, S.~Ramos-Sanchez, M.~Ratz and P.~K.~S.~Vaudrevange,
  %``(Non-)Abelian discrete anomalies,''
Nucl.\ Phys.\ B {\bf 805}, 124 (2008)
%doi:10.1016/j.nuclphysb.2008.07.005
[arXiv:0805.0207 [hep-th]].
%%CITATION = doi:10.1016/j.nuclphysb.2008.07.005;%%  

%\cite{Ishimori:2010au}
\bibitem{Ishimori:2010au} 
  H.~Ishimori, T.~Kobayashi, H.~Ohki, Y.~Shimizu, H.~Okada and M.~Tanimoto,
  %``Non-Abelian Discrete Symmetries in Particle Physics,''
Prog.\ Theor.\ Phys.\ Suppl.\  {\bf 183}, 1 (2010)
%doi:10.1143/PTPS.183.1
[arXiv:1003.3552 [hep-th]].
%%CITATION = doi:10.1143/PTPS.183.1;%%  
  
  
 %\cite{Blumenhagen:2006xt}
\bibitem{Blumenhagen:2006xt} 
  R.~Blumenhagen, M.~Cvetic and T.~Weigand,
  %``Spacetime instanton corrections in 4D string vacua: The Seesaw mechanism for D-Brane models,''
Nucl.\ Phys.\ B {\bf 771}, 113 (2007)
%doi:10.1016/j.nuclphysb.2007.02.016
[hep-th/0609191].
%%CITATION = doi:10.1016/j.nuclphysb.2007.02.016;%% 
  
%\cite{Ibanez:2006da}
\bibitem{Ibanez:2006da} 
  L.~E.~Ibanez and A.~M.~Uranga,
  %``Neutrino Majorana Masses from String Theory Instanton Effects,''
JHEP {\bf 0703}, 052 (2007)
%doi:10.1088/1126-6708/2007/03/052
[hep-th/0609213].
%%CITATION = doi:10.1088/1126-6708/2007/03/052;%%  
  
%\cite{Cvetic:2007ku}
\bibitem{Cvetic:2007ku} 
  M.~Cvetic, R.~Richter and T.~Weigand,
  %``Computation of D-brane instanton induced superpotential couplings: Majorana masses from string theory,''
Phys.\ Rev.\ D {\bf 76}, 086002 (2007)
doi:10.1103/PhysRevD.76.086002
[hep-th/0703028].
%%CITATION = doi:10.1103/PhysRevD.76.086002;%%  
  
%\cite{Kobayashi:2015siy}
\bibitem{Kobayashi:2015siy} 
  T.~Kobayashi, Y.~Tatsuta and S.~Uemura,
  %``Majorana neutrino mass structure induced by rigid instantons on toroidal orbifold,''
Phys.\ Rev.\ D {\bf 93}, no. 6, 065029 (2016)
%doi:10.1103/PhysRevD.93.065029
[arXiv:1511.09256 [hep-ph]]; 
%%CITATION = doi:10.1103/PhysRevD.93.065029;%%  
%\cite{Kobayashi:2016ovu}
%\bibitem{Kobayashi:2016ovu} 
  T.~Kobayashi, S.~Nagamoto and S.~Uemura,
  %``Modular symmetry in magnetized/intersecting D-brane models,''
arXiv:1608.06129 [hep-th].
%%CITATION = ARXIV:1608.06129;%%

%\cite{Sumita:2015sta}
\bibitem{Sumita:2015sta}
  K.~Sumita,
  %``Quasi-localized wavefunctions on magnetized tori and tiny neutrino Yukawa couplings,''
  JHEP {\bf 1601} (2016) 067
  %doi:10.1007/JHEP01(2016)067
  [arXiv:1509.03392 [hep-ph]].
  %%CITATION = doi:10.1007/JHEP01(2016)067;%%

%\cite{Kobayashi:1973fv}
\bibitem{Kobayashi:1973fv}
  M.~Kobayashi and T.~Maskawa,
  %``CP Violation in the Renormalizable Theory of Weak Interaction,''
  Prog.\ Theor.\ Phys.\  {\bf 49} (1973) 652.
  %doi:10.1143/PTP.49.652
  %%CITATION = doi:10.1143/PTP.49.652;%%
  %8764 citations counted in INSPIRE as of 06 Sep 2016

%\cite{Pontecorvo:1967fh}
\bibitem{Pontecorvo:1967fh}
  B.~Pontecorvo,
  %``Neutrino Experiments and the Problem of Conservation of Leptonic Charge,''
  Sov.\ Phys.\ JETP {\bf 26} (1968) 984
   [Zh.\ Eksp.\ Teor.\ Fiz.\  {\bf 53} (1967) 1717]; 
  %%CITATION = SPHJA,26,984;%%
  %1614 citations counted in INSPIRE as of 06 Sep 2016
%\cite{Maki:1962mu}
%\bibitem{Maki:1962mu}
  Z.~Maki, M.~Nakagawa and S.~Sakata,
  %``Remarks on the unified model of elementary particles,''
  Prog.\ Theor.\ Phys.\  {\bf 28} (1962) 870.
  %doi:10.1143/PTP.28.870
  %%CITATION = doi:10.1143/PTP.28.870;%%
  %3024 citations counted in INSPIRE as of 06 Sep 2016

%\cite{Beringer:1900zz}
\bibitem{Beringer:1900zz}
  J.~Beringer {\it et al.}  [Particle Data Group Collaboration],
  %``Review of Particle Physics (RPP),''
  Phys.\ Rev.\ D {\bf 86} (2012) 010001.
  %%CITATION = PHRVA,D86,010001;%%

%\cite{Sasaki:1986jv}
\bibitem{Sasaki:1986jv}
  K.~Sasaki,
  %``Renormalization Group Equations for the {Kobayashi-Maskawa} Matrix,''
  Z.\ Phys.\ C {\bf 32} (1986) 149.
  %doi:10.1007/BF01441364
  %%CITATION = doi:10.1007/BF01441364;%%
  %56 citations counted in INSPIRE as of 02 Sep 2016

%\cite{Kobayashi:2016qag}
\bibitem{Kobayashi:2016qag}
  T.~Kobayashi, K.~Nishiwaki and Y.~Tatsuta,
  %``CP-violating phase on magnetized toroidal orbifolds,''
  arXiv:1609.08608 [hep-th].
  %%CITATION = ARXIV:1609.08608;%%

%\cite{Choi:2009pv}
\bibitem{Choi:2009pv}
  K.~S.~Choi, T.~Kobayashi, R.~Maruyama, M.~Murata, Y.~Nakai, H.~Ohki and M.~Sakai,
  %``E(6,7,8) Magnetized Extra Dimensional Models,''
  Eur.\ Phys.\ J.\ C {\bf 67} (2010) 273
 % doi:10.1140/epjc/s10052-010-1275-9
  [arXiv:0908.0395 [hep-ph]];
  %%CITATION = doi:10.1140/epjc/s10052-010-1275-9;%%
  %21 citations counted in INSPIRE as of 23 Oct 2016
%
%\cite{Kobayashi:2010an}
%\bibitem{Kobayashi:2010an}
  T.~Kobayashi, R.~Maruyama, M.~Murata, H.~Ohki and M.~Sakai,
  %``Three-generation Models from E_8 Magnetized Extra Dimensional Theory,''
  JHEP {\bf 1005} (2010) 050
 % doi:10.1007/JHEP05(2010)050
  [arXiv:1002.2828 [hep-ph]];
  %%CITATION = doi:10.1007/JHEP05(2010)050;%%
  %15 citations counted in INSPIRE as of 23 Oct 2016
%
%\cite{Abe:2015mua}
%\bibitem{Abe:2015mua}
  H.~Abe, T.~Kobayashi, H.~Otsuka and Y.~Takano,
  %``Realistic three-generation models from SO(32) heterotic string theory,''
  JHEP {\bf 1509} (2015) 056
 % doi:10.1007/JHEP09(2015)056
  [arXiv:1503.06770 [hep-th]].
  %%CITATION = doi:10.1007/JHEP09(2015)056;%%
  %6 citations counted in INSPIRE as of 23 Oct 2016

%\cite{Abe:2016zgq}
\bibitem{Abe:2016zgq}
  H.~Abe, T.~Kobayashi and K.~Sumita,
  %``Dynamical supersymmetry breaking on magnetized tori and orbifolds,''
  Nucl.\ Phys.\ B {\bf 911} (2016) 606
 % doi:10.1016/j.nuclphysb.2016.08.030
  [arXiv:1605.02922 [hep-th]].
  %%CITATION = doi:10.1016/j.nuclphysb.2016.08.030;%%

%\cite{Lee:2003mc}
\bibitem{Lee:2003mc}
  H.~M.~Lee, H.~P.~Nilles and M.~Zucker,
  %``Spontaneous localization of bulk fields: The Six-dimensional case,''
  Nucl.\ Phys.\ B {\bf 680} (2004) 177
 % doi:10.1016/j.nuclphysb.2003.12.031
  [hep-th/0309195].
  %%CITATION = doi:10.1016/j.nuclphysb.2003.12.031;%%
  %46 citations counted in INSPIRE as of 23 Oct 2016

%\cite{GrootNibbelink:2002wv}
\bibitem{GrootNibbelink:2002wv}
  S.~Groot Nibbelink, H.~P.~Nilles and M.~Olechowski,
  %``Spontaneous localization of bulk matter fields,''
  Phys.\ Lett.\ B {\bf 536} (2002) 270
 % doi:10.1016/S0370-2693(02)01840-3
  [hep-th/0203055];
  %%CITATION = doi:10.1016/S0370-2693(02)01840-3;%%
  %60 citations counted in INSPIRE as of 23 Oct 2016
%
%\cite{GrootNibbelink:2002qp}
%\bibitem{GrootNibbelink:2002qp}
%  S.~Groot Nibbelink, H.~P.~Nilles and M.~Olechowski,
  %``Instabilities of bulk fields and anomalies on orbifolds,''
  Nucl.\ Phys.\ B {\bf 640} (2002) 171
 % doi:10.1016/S0550-3213(02)00564-3
  [hep-th/0205012];
  %%CITATION = doi:10.1016/S0550-3213(02)00564-3;%%
  %71 citations counted in INSPIRE as of 23 Oct 2016
%
%\cite{Abe:2002ps}
%\bibitem{Abe:2002ps}
  H.~Abe, T.~Higaki and T.~Kobayashi,
  %``Wave function profile and SUSY breaking in 5-D model with Fayet-Iliopoulos term,''
  Prog.\ Theor.\ Phys.\  {\bf 109} (2003) 809
 % doi:10.1143/PTP.109.809
  [hep-th/0210025].
  %%CITATION = doi:10.1143/PTP.109.809;%%
  %20 citations counted in INSPIRE as of 23 Oct 2016

%\cite{Abe:2004yk}
\bibitem{Abe:2004yk}
  H.~Abe, K.~Choi and I.~W.~Kim,
  %``Fayet-Iliopoulos terms in 5-D orbifold supergravity,''
  JHEP {\bf 0409} (2004) 001
 % doi:10.1088/1126-6708/2004/09/001
  [hep-th/0405100];
  %%CITATION = doi:10.1088/1126-6708/2004/09/001;%%
  %14 citations counted in INSPIRE as of 23 Oct 2016
%
%\cite{Correia:2004pz}
%\bibitem{Correia:2004pz}
  F.~Paccetti Correia, M.~G.~Schmidt and Z.~Tavartkiladze,
  %``(BPS) Fayet-Iliopoulos terms in 5-D orbifold SUGRA,''
  Phys.\ Lett.\ B {\bf 613} (2005) 83
 % doi:10.1016/j.physletb.2005.03.002
  [hep-th/0410281];
  %%CITATION = doi:10.1016/j.physletb.2005.03.002;%%
  %17 citations counted in INSPIRE as of 23 Oct 2016
%
%\cite{Abe:2004nx}
%\bibitem{Abe:2004nx}
  H.~Abe and K.~Choi,
  %``Gauged U(1)(R) symmetries and Fayet-Iliopoulos terms in 5-D orbifold supergravity,''
  JHEP {\bf 0412} (2004) 069
 % doi:10.1088/1126-6708/2004/12/069
  [hep-th/0412174].
  %%CITATION = doi:10.1088/1126-6708/2004/12/069;%%
  %10 citations counted in INSPIRE as of 23 Oct 2016

\end{thebibliography}
\end{document}